# FUNCTIONALIZATION OF SINGLE-WALLED CARBON NANOTUBES WITH BIS-ASTRAPHLOXIN

EBRIMA SAHO

APPLIED PHYSICS BSC

2019

ASTON UNIVERSITY

# Summary

This dissertation aims to demonstrate that photoluminescence spectroscopy could be an efficient technique of functionalized carbon nanotube detection and imaging. The advances in nanotechnology and the potential growth in carbon nanotubes production, in which if control mechanisms not in place for safely disposing off nanotubes, these nanotubes could pose as an environmental pollutant, due to the severe, persistent and bio-accumulative nature of these compounds. Therefore, the need for this scientific research is timely in case of some technological disaster. Besides, the study on interaction of carbon nanotubes with various organic materials provide a new knowledge and opportunities for potential application of the nanotubes.

The dissertation starts by exploring and conducting an in-depth literature survey on the functionalization of carbon nanotubes, energy transfer and photoluminescence imaging and the project develops upon the previous work by P. Lutsyk et al. in Light Sci Appl. 5 (2016) e16028 entitled *"A Sensing Mechanism for the Detection of Carbon Nanotubes Using Selective Photoluminescent Probes Based on Ionic Complexes with Organic Dyes."* The literature is followed by comprehensive experimental studies by optical spectroscopy techniques on a new system of carbon nanotubes with bis-astraphloxin, which is promising to provide a new insight into non-covalent functionalization of carbon nanotubes. The bis-astraphloxin is a new cyanine dye molecule consisting of two astraphloxin units linked together, where individual astraphloxin units have been studied before and showed strong energy transfer to the nanotubes.

The main findings from the experimental studies described in this dissertation illustrate that carbon nanotubes interact with bis-astraphloxin. This is evidenced by absorption spectra of the mixtures are not a superposition of the spectra of each component: the nanotubes or bis-astraphloxin dye only. In the studied mixtures of the nanotubes with bis-astraphloxin, we observe a strong quenching of visible range dye emission and significant enhancement (up to a factor of 4) of the near infrared emission of carbon nanotubes in the range of the dye excitation. This result is evidencing efficient energy transfer from the interacting donor, bis-astraphloxin, to acceptor, the nanotubes. Comparison of the bis-astraphloxin-nanotube mixture with the mixture of astraphloxin-nanotubes and relating to literature data, it is possible to associate the interaction and energy transfer due to the formation of ionic complexes of bis-astraphloxin-SWCNT. The compounds are self-assembled due to Coulomb attraction of positively charged bis-astraphloxin and positively charged nanotubes covered by negatively charged surfactant. Finally. it has been shown that higher concentration of bis-astraphloxin contribute to the formation of such complexes, which could be used for further design of functionalized carbon nanotubes.

# Acknowledgments

I am very humbled and blessed to get involved in nanoscience research and now that this project has been completed. It equipped me with invaluable knowledge and expertise in the field of scientific research. Even though the project was highly demanding, I have gained all the necessary tools and skills. For example, it boosted my confidence in problem-solving, communication and I have progressed immensely in understanding empirical analysis of experimental data. It was a significant and exciting experience.

Therefore, I will begin to give thanks and praises to the Almighty God for providing me the strength to venture into this critical exploration.

I want to express my profound appreciation to my primary supervisor, Dr Petro Lutsyk who suggested this exciting research area and gave me all the necessary training before the start of the project. He supported and supervised me throughout this project. His valuable guidance and constant motivation have been the driving force that assisted me to remain focused on this project and mould it into full proof success. His recommendations and instructions were vital in the realization of the project.

Many thanks to my secondary supervisor Dr. Richard Martin for his initial suggestions and for aiding to ensure that my project remains on course.

Also, I would like to cordially thank Prof. Yu. Slominski and Dr. M. Shandura for the dye provision.

Finally, I would like to thank my friends and my family for their tremendous support and advice especially in the last few weeks of this project.

# Contents

# List of Figures

# List of Tables

# Abbreviations

| | |
|---|---|
| SWCNT/CNT | Single-walled-carbon nanotubes/ Carbon nanotubes |
| DWCNT | Double-walled-carbon nanotubes |
| MWCNT | Multi-walled-carbon nanotubes |
| NIR | Near infrared |
| VIS | Visible |
| UV | Ultraviolet |
| PO | Primary objective |
| EO | Extended objective |
| SDS | Sodium dodecyl sulphate |
| SDBS | Sodium dodecylbenzene sulfonate |
| AFM | Atomic force microscopy |
| RET | Resonance energy transfer |
| FRET | Förster resonance energy transfer |
| QY | Quantum yield |
| T | Transmittance |
| CBM | Common beam mask |
| InGaAs | Indium gallium arsenide phosphide |
| $\lambda_{max}$ | absorbance maximum |
| D | donor |
| A | acceptor |
| D/A | donor-acceptor |
| $\lambda_{EX}$ or EX wavelength | excitation wavelength |
| $\lambda_{EM}$ or EM wavelength | emission wavelength |
| PL | photoluminescence |
| PLE map | PL excitation-emission maps |

# 1 Introduction

The discovery of carbon nanotubes captured the attention of researchers due to their combination of desirable intrinsic properties such as high tensile strength, elasticity, electrical and thermal conductivity. The research effort continues due to their excellent configurability and broad scope for modification giving many avenues for investigation and application.

Of special interest are 1-dimensional single-walled carbon nanotubes (SWCNTs) formed by rolling 2-dimensional graphene. This structure benefits from intrinsic near-infrared (NIR) fluorescence (900-1400 nm), which overlaps the spectral region in which biological tissues are optically transparent. Functionalized nanotubes are then ideal for highly targeted detection, monitoring, and therapy of diseases [1]. However, as nanotubes are resilient to natural processes of degradation, they can endure in the environment for long periods of time thus posing a problem as a pollutant potentially disrupting the ecosystem. For this reason, an efficient method for nanotube detection is required.

This project started in September 2018 and ended in May 2019. The project develops upon the work by P. Lutsyk et al. [2] and aims to demonstrate that photoluminescence spectroscopy could be an efficient technique of carbon nanotube detection and imaging. Current issues which this project seeks to address includes:

- The enhancement of photoluminescence from SWCNTs via energy transfer from organic dyes.
- The stability of non-covalently interacting SWCNT-dye systems not fully explored.

Of particular interest are the physical mechanisms for the interaction of carbon nanotubes with the previously non-investigated organic dye, bis-astraphloxin, which is supposed to yield significant amplification of photoluminescence response via resonant energy transfer. The project was initially launched to address the problems via the study of a novel SWCNT-dye system with energy transfer and exploration of a new organic dye, bis-astraphloxin. Bis-astraphloxin is a new cyanine dye molecule consisting of two astraphloxin units linked together forming a modified analogue of astraphloxin that showed a significant enhancement of photoluminescence via energy transfer [2]. In early November 2018 after the project plan and proposal was presented to the secondary supervisor Dr. Richard Martin. The suggestion was that it would be necessary if astraphloxin is also included in the investigation to compare and use it as a reference system. The difference between the two systems is the type of dye used. The new system uses a novel organic dye called bis-astraphloxin as in Figure 1, while the reference system uses simple analogue astraphloxin as in Figure 2.

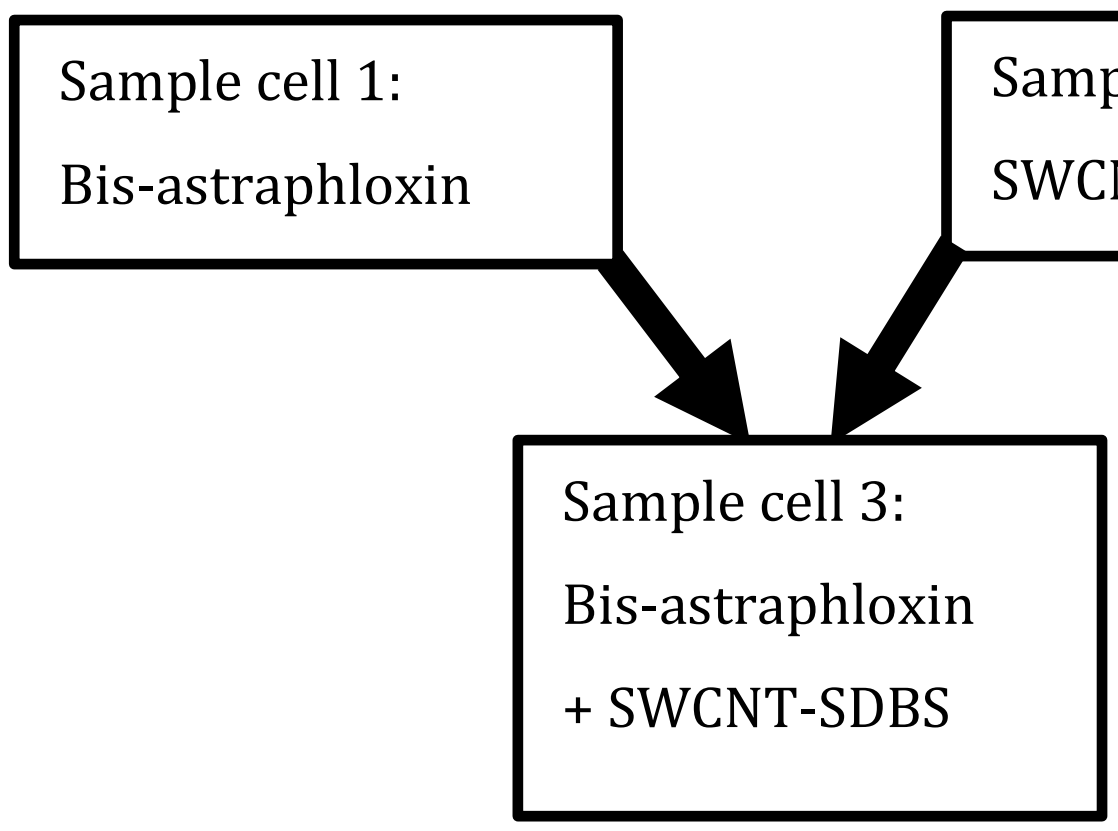

**Figure 1 - New system**

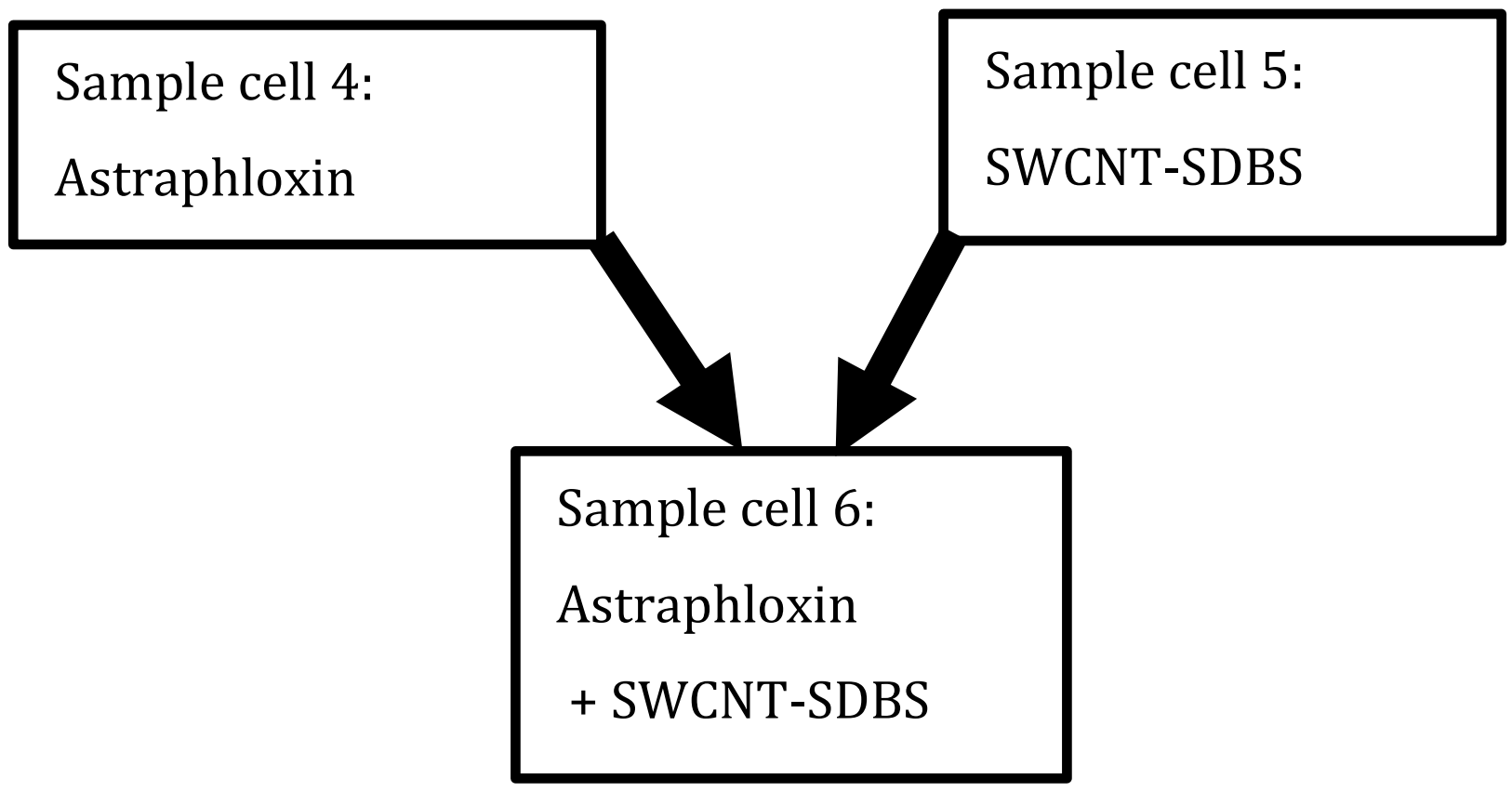

**Figure 2 - Reference system**

This dissertation will focus on the uninvestigated dye called bis-astraphloxin and will briefly look into the previously studied dye named astraphloxin for comparison reasons and to aid in concluding which dye is better at enhancing photoluminescence through resonance energy transfer based on the following primary and extended objectives: -

Primary Objectives:

- Literature review on SWCNT functionalization, energy transfer, and photoluminescence imaging.
- Experiments with mixtures of SWCNTs and bis-astraphloxin to explore energy transfer and enhancement of photoluminescent signal.
- An empirical analysis of experimental data, to develop the mechanism of interaction of bis-astraphloxin with SWCNTs.

Extended Objectives:

- Experiments and analysis of temporal evolution of energy transfer or aggregation in the SWCNT-dye mixtures.

Therefore, for the successful completion of this project, different milestones and deliverables dates were set from September 2018 to May 2019. The following primary and extended objectives show the individual landmark and deliverable end dates.

# 1.1 Primary Objectives (PO) of the Project

**PO1 – Literature Survey on SWCNT Functionalization, Energy Transfer and Photoluminescence Imaging**

**Milestone M1.1** – Preliminary literature review (completed by 15$^{th}$ of October 2018)

**Deliverable D1.1** – Literature review as a chapter for the dissertation (completed by 14$^{th}$ of December 2018).

**PO2 – Experiment for Mixtures of SWCNT with Bis-Astraphloxin**

**Milestone M2.1** – Preliminary training on experimental measurements of absorption and photoluminescence spectra, including excitation-emission photoluminescence maps (completed by 25$^{th}$ of January 2019).

**Milestone M2.2** – Samples preparation: neat dispersion of SWCNT, neat solution of the dye, and mixtures of the SWCNT with bis-astraphloxin (completed by 18$^{th}$ of February 2019).

**Deliverable D2.1** – Measurements of absorption and photoluminescence spectra for the samples (completed by 18$^{th}$ of February 2019).

**PO3 – Empirical Analysis of Experimental Data**

**Deliverable D3.1** – Physical mechanism proposed for the energy transfer in the SWCNT-dye system (completed by 22$^{nd}$ of February 2019).

# 1.2 Extended Objective (EO) of the Project

**EO4 – Experiment and Analysis of Temporal Developments of Energy Transfer/Aggregation in the SWCNT-Dye Mixtures**

**Deliverable D4.1** – Temporal measurements of absorption and photoluminescence spectra for the samples (completed by 11$^{th}$ of March 2019).

**Deliverable D4.2** – An updated physical mechanism for the SWCNT-dye energy transfer (completed by 22$^{nd}$ of March 2019)

For the purpose of data acquisition and for effective empirical data analysis, OriginLab was selected because of the various in-built tools for linear, polynomial and nonlinear curve and surface fitting. It also provides several features for peak analysis, which is very helpful throughout the course of this project.

# 2 Background

The inspiration for this project is the extension upon the advances in functionalization of carbon nanotubes and its practical applications in reinforcement composites, biomedicine, sensors, and energy generation/transport/storage [1-3]. For example, unique electrical, thermal and spectral properties of carbon nanotubes functionalized by organic molecules offer further biomedical advances in the detection, monitoring, and therapy of diseases because the cells can comfortably internalize the functionalized carbon nanotubes and consequently acts as a transportation vehicle for different molecules applicable to treatment and diagnosis [3,4].

The energy transfer from the organic molecules to SWCNTs enables robust sensitization of photoluminescent response from the nanotubes and this way the SWCNTs detected in the natural environment in case of some technological disaster [2]. The literature review of SWCNT functionalization, energy transfer, and photoluminescence imaging cover 10 main topics which this dissertation explored. Mentioned below are the subjects discretely highlighted.

- The Allotropes Carbon
- Carbon nanotubes
- Purification of carbon nanotubes
- Molecular fluorescence
- Resonance energy transfer
- Photoluminescence imaging
- UV/VIS/NIR Spectroscopy
- Dye (bis-astraphloxin)
- Dye (astraphloxin)
- Surfactant (SDBS)

This literature review studied each basic theme concepts and practical applications. Therefore, the following chapter will discuss the literature survey on the following topics above in detail.

## 2.1 The Allotropes of Carbon

Carbon is an element with atomic number 6 that can be found on the periodic table (see Figure 3). It provides the basis for life on earth. Elemental carbon has many technological applications, ranging from drugs to synthetic materials. Due to the structure and electronic configuration of Carbon, it can bind to itself and practically to most of the other elements in an endless variety. The resulting structural diversity of organic compounds and molecules accompanied by a broad range of chemical and physical properties [4].

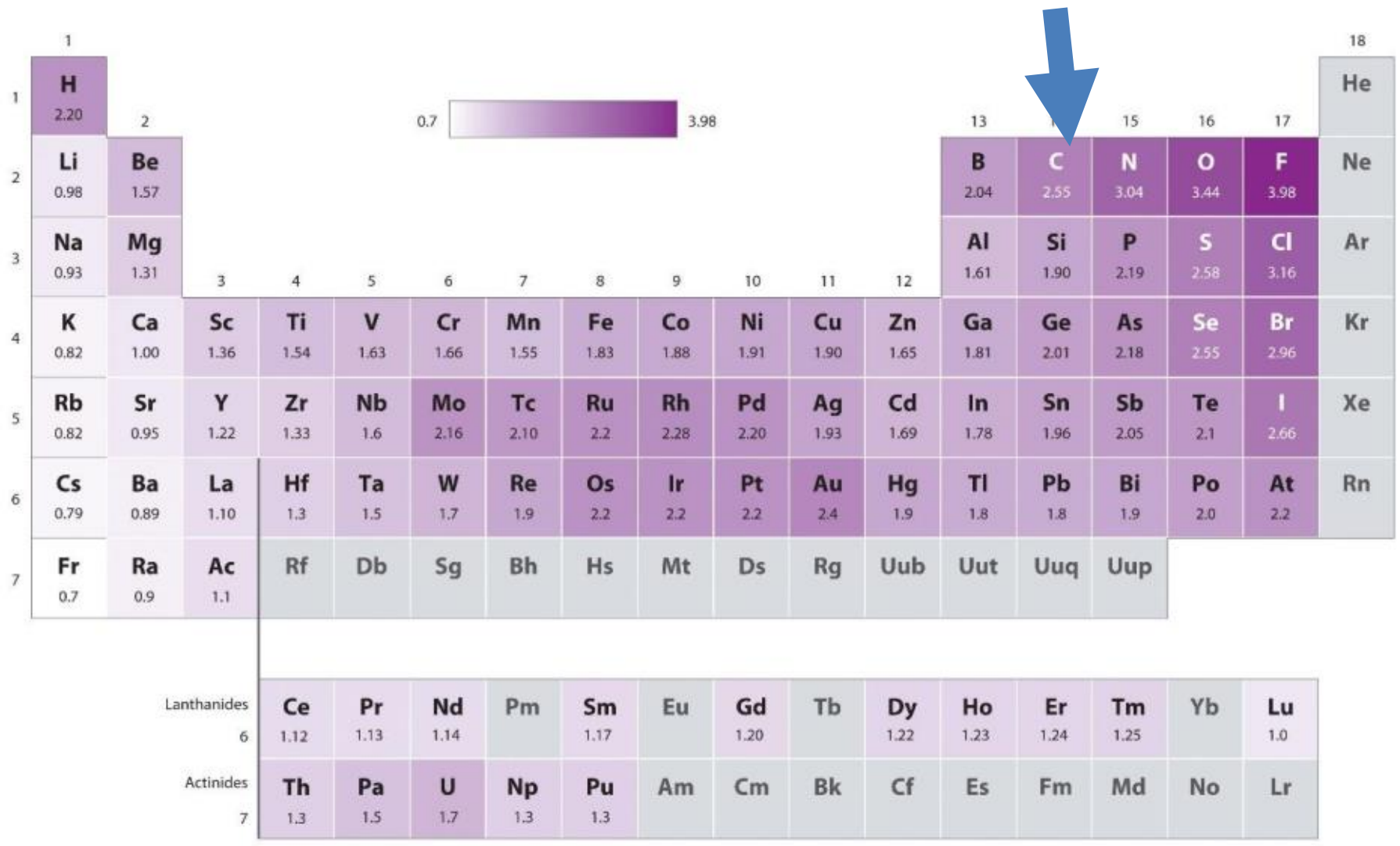

**Figure 3: The periodic table indicating carbon [5].**

The principal carbon as shown in the periodic table exists in two natural allotropes, diamond, and graphite, which consist of extended networks of sp3- and sp2-hybridized carbon atoms, respectively. These natural allotropes indicated distinctive physical properties such as hardness, thermal conductivity, lubrication behaviour or electrical conductivity. The allotropes can be constructed in many ways by altering the periodic binding shape in networks consisting of sp3-, sp2- and sp-hybridized carbon atoms [4].

The two known allotropes of carbon existed for a while before the discovery of fullerene in 1985, carbon nanotubes in 1991, and graphene in 2004 as shown in Figure 4. The observations marked the beginning of an ***era of synthetic carbon allotropes***. The countless possible ways of alterations and number of scientists probing this challenge, guarantee that the revelations and discoveries in this area will continue [4].

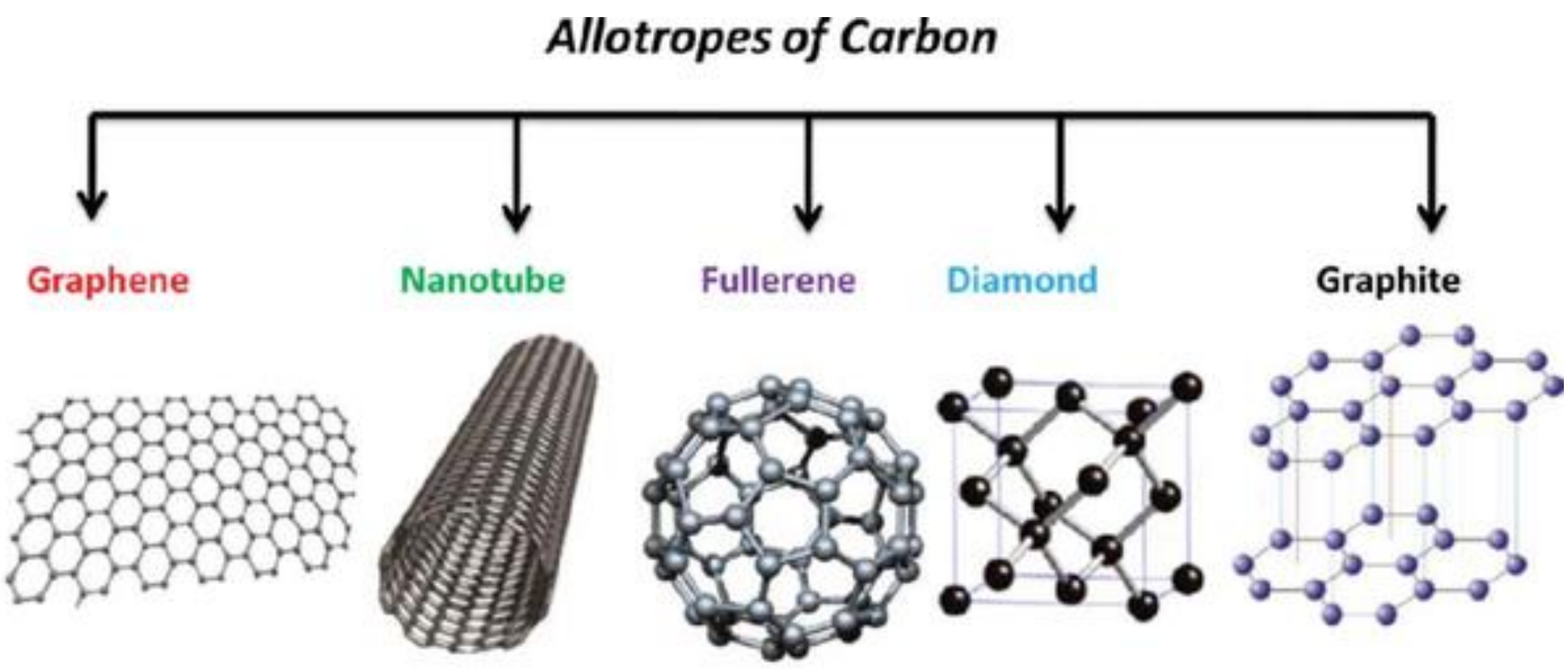

**Figure 4: The generation of synthetic carbon allotropes [6].**

# 2.2 Carbon Nanotubes

The discovery of a spherical molecule composed exclusively of carbon atoms as in Figure 5 was uncovered in 1996 by three great scientist namely Harry Kroto, Robert Curl, and Richard Smalley. These scientists were awarded a Nobel prize in Chemistry for such a phenomenal revelation [7].

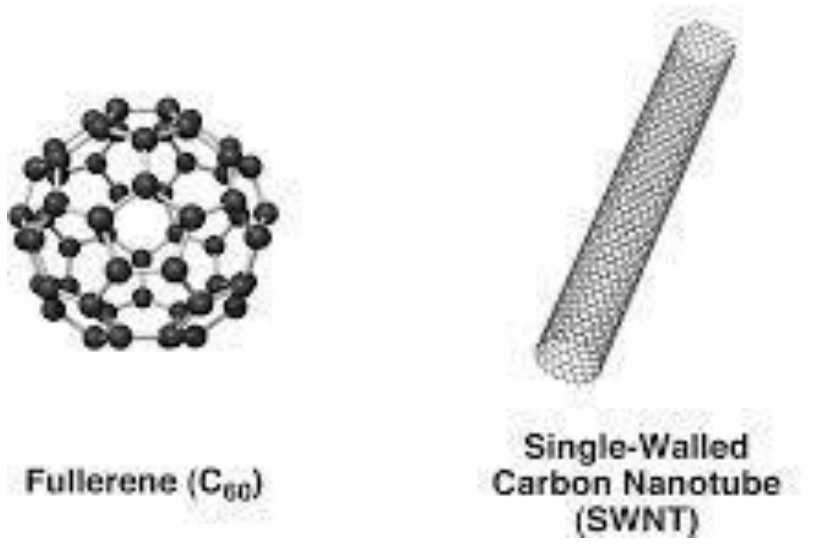

**Figure 5: Fullerene C60 model on the left [8]. Fullerenes are close cage clusters relatively stable in the gas phase [7]. The single-walled carbon nanotubes on the right are elongated structures that can reach several microns in length and small diameter in the nanometer range.**

The nanometre-scale framework was named “fullerene” because of its resemblance to the extremely conformity architectonic geodesic domes designed by the architect Richard Buckminster Fuller. Extensive studies were carried out in the 1980s and the early 1990s on fullerene theory, synthesis, and its characterization. In 1991 Lijima, presented transmission electron microscopy observations of elongated and concentric layered microtubules made of carbon atoms, which considered as filamentous carbon. The surveillance by Lijima, catapulted the substantial investigations corresponding to one of the most actively studied shapes of the last century, presently named carbon nanotubes (CNTs) [7].

## 2.2.1 Types of Carbon Nanotubes and Related Structures

There are different types of carbon nanotubes, which are given below. The focus here is the single-walled carbon nanotubes, and other types of carbon nanotubes not discussed in this review.

- Single-walled carbon nanotubes (SWCNTs)
- Multi-walled carbon nanotubes (MWCNTs)
- Double -walled carbon nanotubes (DWCNTs)
- Junctions and Crosslinking
- Other morphologies
- Extreme carbon nanotubes

### 2.2.2 Single-Walled Carbon Nanotubes

The shape and structure of carbon nanotubes conceptualized as a rolled-up sheet of graphene (Figure 6), which is a planar – hexagonal arrangement of carbon atoms distributed in a honeycomb lattice [7].

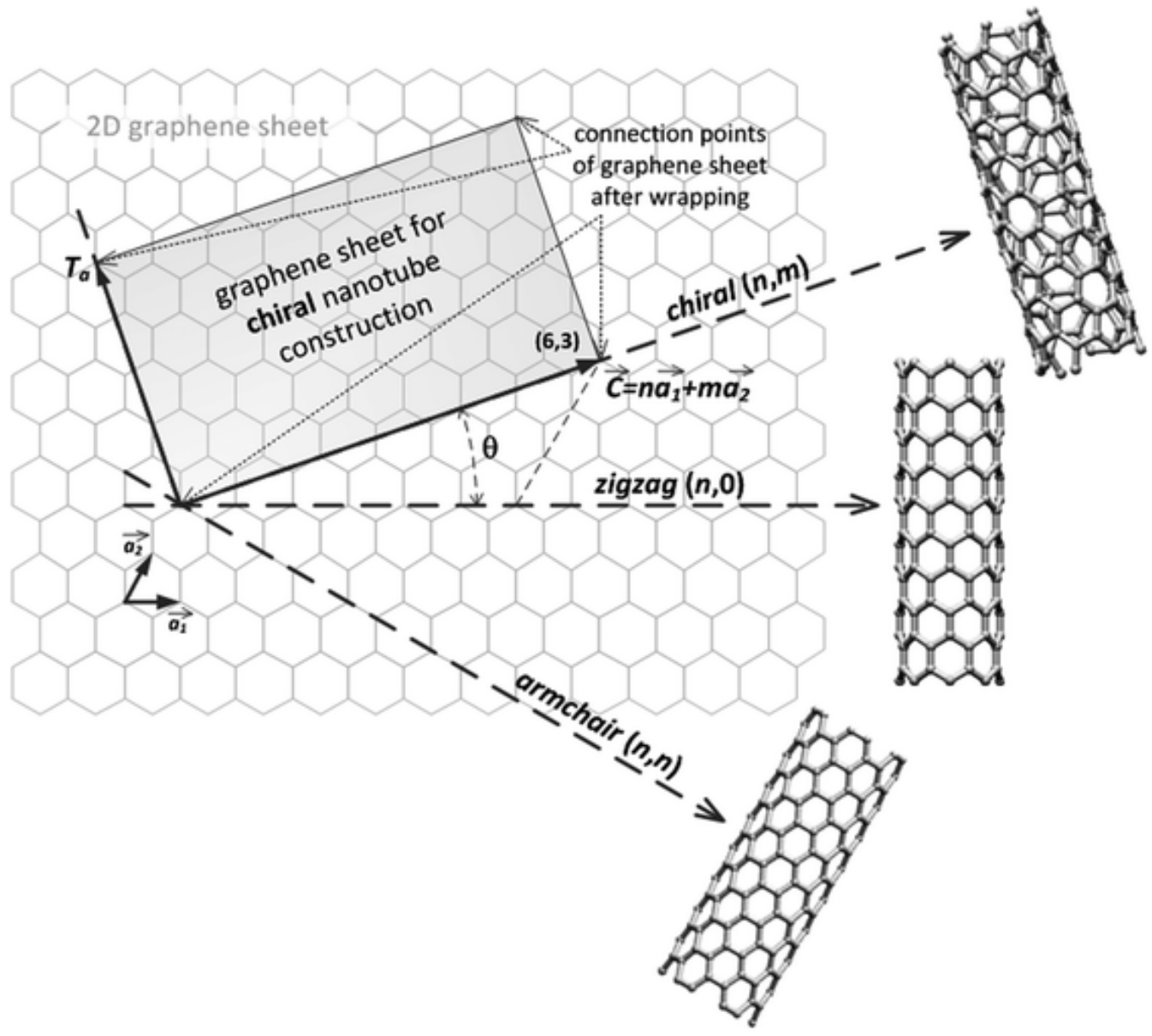

**Figure 6: A single-walled carbon nanotubes formed from 2-dimensional graphene [9].**

From Figure 6 we can see that the nanotubes are classified into three main categories according to the rolling direction, namely, zigzag, armchair, and chiral. From an underlying perspective and for future practical applications the most prominent characteristics of SWCNT are that they exhibit unique electronic (semiconducting or metallic), mechanical (Young Modulus), optical and chemical characteristics [4].

### 2.2.3 Carbon Nanotubes Production

There are different methods for producing SWCNT as indicated below. Each of the production techniques has an associated advantage as well as disadvantage. The most established high-temperature techniques are arc discharge and laser ablation, as well as chemical vapour deposition (CVD) with its common variants [4]. The literature regarding the synthesis of SWCNTs is inexhaustible, but far from being fully understood.

- Arc discharge
- Laser ablation
- Chemical vapour deposition
- Miscellaneous synthesis methods

The state-of-the-art carbon nanotubes production entails many approaches and new paths continuously evolved.

### 2.2.4 Functionalization of Carbon Nanotubes

Functionalization of carbon nanotubes have two broad categories (see Figure 7)

- covalent functionalization
- non-covalent functionalization

The idea of functionalizing carbon nanotubes implemented because of the outer wall of the pristine nanotubes being resistant and conceived as chemically inert. The inactive behaviour of these nanotubes is not consistently advantageous for applications and this ground, modifying the properties of carbon nanotubes in a controlled manner have been thought of to make them chemically active. The process of chemically activating can be implemented using one of the several functionalization routes convenient. It is crucial revealing that in many cases the term functionalization could be too broad to explain the kind of changes desired. The concept means in-lattice doping, intercalation, molecule or particle adsorption, encapsulation, or even other nonexposed modifications [4].

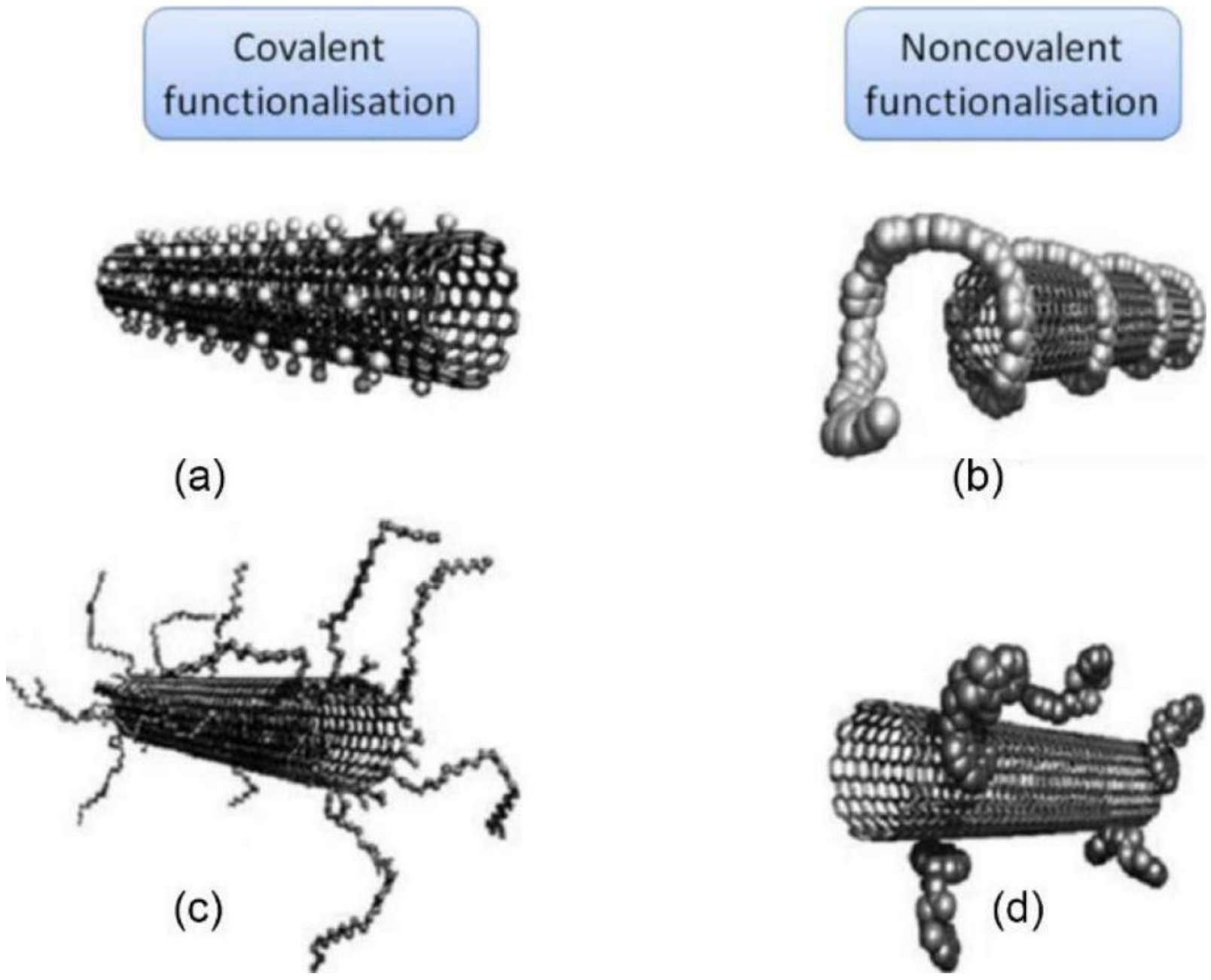

**Figure 7: Different routes for nanotubes functionalization: (a) sidewall covalent functionalization, (b) defect-group covalent functionalization, (c) Non- covalent polymer wrapping, (d) Non-covalent pi-stacking.**

The centre of attention in the next subsection will be non-covalent functionalization of carbon nanotubes, which is the focal point in this research. This survey will not explore the covalent functionalization of carbon nanotubes.

### 2.2.5 Non-Covalent Functionalization of Carbon Nanotubes

CNTs manifested unique mechanical, optical and electrical properties that model them perfect nanoscale materials. Notwithstanding, the suggested application mentioned above limited by their experimental insolubility in aqueous and organic solvents. Due to the high polarizability and smooth surface of carbon nanotubes, and, SWCNTs in particular – form bundles and ropes, whereby individual nanotubes are positioned parallel to each other with Van der Waals attraction. Also, CNTs obtained as mixtures that demonstrated different chiralities, diameter, and length, in which non-carbon nanotubes and metal catalysts are also present in the final material [7].

Some of these hindrances overpowered by the controlled defect and sidewall functionalization of CNTs. The formation of covalent linkages can drastically enhance the solubility of CNTs in various solvents while guarantees the structural integrity of the nanotube skeleton, but it also alters the intrinsic physical properties of the CNTs due to the modification of the sp2 carbon framework. The most important effect is that the inherent conductivity of the CNTs destroyed. [7]

A possible and alternative strategy for preserving the unique electronic and mechanical properties of CNTs is by non-covalent or supramolecular modifications of CNTs. Non-covalent interactions entail hydrophobic, van der Waals, and electrostatic forces, and require the physical adsorption of suitable molecules onto the sidewalls of the CNTs. Non-covalent functionalization is achieved by polymer wrapping, adsorption of surfactants or small aromatic molecules, and interaction with porphyrins or biomolecules such as DNA and peptides [7].

### 2.2.6 Non-Covalent Interactions of CNTs with Surfactants and Ionic Liquids

Significant progress towards the solubilization of CNTs in water, which is crucial because of potential biomedical applications and biophysical processing schemes, has been achieved by using surfactants. Among the surfactants, the anionic surfactant sodium dodecyl sulphate (SDS) has been the most widely used surfactant [7].

Surfactant molecules containing aromatic groups can form more specific and more direction π-π stacking interactions with the graphite surface of CNTs. Interactions of SDS and the structurally related sodium dodecylbenzene sulfonate (SDBS) with SWCNTs have been compared to indicate the role of the aromatic rings. SDS and SDBS have the same length of the alkyl chain, but the latter has a phenyl ring attached between the alkyl chain and the hydrophilic group. The presence of the phenyl ring makes SDBS more effective for solubilization of CNTs than SDS, because of the aromatic stacking formed between the SWCNTs and the phenyl rings of the SDBS within the micelle. The diameter distribution of CNTs in the dispersion of SDBS, measured by atomic force microscopy (AFM), showed that even at 20mg/ml ~63 ± 5% of the SWCNT bundles exfoliate in single tubes [7].

# 2.3 Purification of Carbon Nanotubes

CNT material decontaminated due to the different procedures of production employed. The methods of synthesis utilized vary from each other, and the presence of unwanted by-products in the soot is unavoidable. These contaminants interfere with the exceptional properties of the nanotubes. The method of determining the degree of nanotubes purity is debatable because there is no consolidated standard to follow based on the reference uses of synthesized materials. Several methods used in SWCNT purification and none of the mechanisms is ubiquitous. The cleaning employs different oxidation steps, and the main processes given below.

- Sonication
- Thermal treatments
- Chemical acid treatments
- Others

It is very paramount to understand even after the processes mentioned above; nanotubes still need to be separated because of the existence of a different number of walls in the mixture of the nanotubes. There are 2 well known possible separation techniques of nanotubes called electrophori-based and ultracentrifugation-based. However, the idea of purifying a nanotube is still questionable due to the number of parameters that could be considered [7].

# 2.4 Molecular Fluorescence

## 2.4.1 Luminescence

The spontaneous emission of radiation from an electronically or vibrationally excited species that are not in thermal equilibrium with its environment [10]. Light emission classified into nine different types, and this grouping based on the mode of excitation. Table 1 gives the various types of fluorescence.

### 2.4.2 Types of Luminescence

**Table 1: Various types of luminescence [13].**

| Phenomenon | Mode of excitation |
|---|---|
| Photoluminescence (fluorescence, phosphorescence, delayed fluorescence) | Absorption of light (photons) |
| Radioluminescence | Ionizing radiation (x-rays, α, β, ϒ) |
| Cathodoluminescence | Cathode rays (electron beam) |
| Electroluminescence | Electric field |
| Thermoluminescence | Heating after prior storage of energy (e.g., radioactive irradiation) |
| Chemiluminescence | Chemical reaction (e.g. oxidation) |
| Bioluminescence | In vivo biochemical reaction |
| Triboluminescence | Frictional and Electrostatic forces |
| Sonoluminescence | Ultrasound |

### 2.4.3 Luminescence Compounds

There are three different types of luminescent compounds

- Organic compounds (aromatic hydrocarbons and derivatives, dyes, etc.)
- Inorganic compounds (doped glasses, semiconductor nanocrystals, metal clusters, carbon nanotubes, and some fullerenes, etc.)
- Organometallic compounds (porphyrin metal complexes, ruthenium complexes, copper complexes, etc.)

Importantly, photoluminescence can be divided into two significant classes called fluorescence or phosphorescence and it is the concept of light absorption by matter, and the absorbing species elevated to an electronically excited state. The spontaneous emission of photons accompanying de-excitation is term as photoluminescence, which is the possible effects resulting from the interaction of light with matter [10]. Stimulated emission is also possible under certain conditions.

### 2.4.4 History of Fluorescence and Phosphorescence

The term phosphorescence comes from 2 Greek words. The first one being 'φωG' meaning light or photon and the second one 'φορεlν' meaning to bear. Therefore, phosphor means something that takes light and is the term assigned to materials that carry glow in the dark after being exposed to light since the middle ages [10].

On the other hand, fluorescence is not that obvious. The idea came from a phenomenal physicist and a professor of mathematics Sir George Gabriel Stokes of Cambridge University in the middle of the $19^{th}$ century. But the primary purpose of fluorescence came from a Spanish physician, Nicolas Monardes, in 1565. He initially observed the incredible blue colour of an infusion of a wood brought from Mexico used to treat Kidney and urinary diseases [10].

# 2.5 Resonance Energy Transfer

The initial consideration of the nonradiative transfer of excitation energy – also called resonance energy transfer (RET) - from a new species to another one was outlined with atoms in the gas phase. G. Cario and J. Franck indicated in 1922 that upon selective excitation of mercury atoms at 254nm in a vapor mixture with thallium atoms, sensitized emission of the latter can be detected at 535nm. H. Kallman and F. London developed a quantum theory of resonance energy transfer via dipole-dipole interaction in the gas phase in 1928. The concept of the critical radius (distance at which transfer, and natural decay of the excited donor are equally probable) was introduced [10].

In solutions, when increasing the concentration of fluorescence in a viscous solvent, E. Gaviola and P. Pringsheim observed in 1924 that the fluorescence polarization gradually decreases but did not explain the result. It was only in 1929 that Francis Perrin correctly explained it because of homo-transfer.  Years before, in 1925, his father Jean Perrin proposed the mechanism of resonance energy transfer. F. Perrin developed in 1932 a quantum mechanical theory of homo-transfer and qualitatively discussed the effect of spectral overlap [10].

A crucial milestone in the chronicle of fluorescence was when Theodor Förster around 1946 developed the complete theory of resonance energy transfer via dipole-dipole interaction and based on the idea of both classical and quantum mechanical approaches.

Instead of resonance energy transfer, the word fluorescence resonance energy transfer was introduced and initially emerged in life science papers. The confusion is because fluorescence does not transition resonance energy transfer, which is a nonradiative process. To avoid the misnomer, it is crucial to consider the F in FRET to stand for Förster or Förster-type instead of fluorescence. Notwithstanding, resonance energy transfer is not just limited to Förster-type transmission, that is, via dipole-dipole interaction.

In the late 1970s and early 1980s, (F) RET has been used as a spectroscopic ruler. It is ideal for measuring the distance between a donor chromophore and an acceptor chromophore in the range of 1 – 10 nm, which used in monitoring or separation of two species [10].

### 2.5.1 Fluorescence Resonance Energy Transfer Process

FRET is a non-radiative procedure whereby a donor (usually a fluorophore) upon excitation goes into an excited state. The excited state transfers energy to a proximal ground state acceptor A through long-range dipole-dipole interactions as in Figure 8. The acceptor absorbs light at the emission wavelength of the donor and does not always have to transmit the light fluorescently (i.e., dark quenching). The rate at which energy transfer occurs mainly depends on factors such as spectral overlap, the relative orientation of the transition dipoles and the distance r between donor and acceptor molecules.

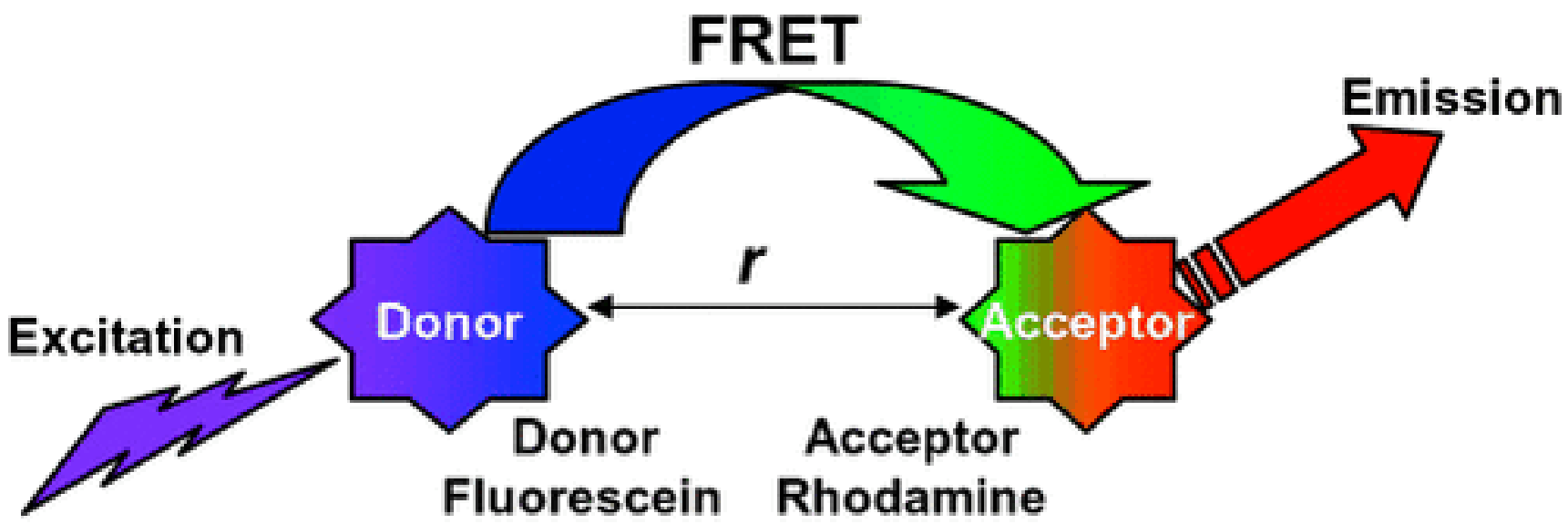

**Figure 8: Schematic illustration of the FRET process: when the donor is excited, the excited molecule transfer energy to the acceptor molecule non-radiatively. A distance r separates the donor and acceptor. The acceptor molecule can release energy either radiatively or non-radiatively [10].**

FRET usually occurs over distances comparable to the dimensions of most biological macromolecules, that is, about 10 to 100 Å. However, formations in which several donors and acceptors interact are progressively typical. The following equations evaluate energy transfer between a single linked D/A pair separated by a fixed distance r which comes from the theoretical treatment of Foster. The energy transfer rate $K_T$(r) between a single D/A pair is dependent on the distance between D and A and can be expressed in terms of the Förster distance $R_0$. $R_0$ is the distance between D and A at which 50% of the excited D molecules decay by energy transfer, while the other half decay through other radiative or non-radiative channels. $R_0$ can be calculated from the spectral properties of the D and A species as in equation 1 [11].

$$R_0 = 9.78 \times 10^3 [K^2 n^{-4} Q_D \mathrm{J}(\lambda)]^{\frac{1}{6}}\ (\text{Å}) \quad \text{(Equation 1)}$$

The factor $K^2$ describes the D/A transition dipole orientation and can range in value 0 (perpendicular) to 4 (collinear/parallel). The controversy about which dipole orientation to assign to a specific FRET format is huge. Only in few cases can the crystal structure of the D/A molecule be determined. There is no other reliable experimental method to measure absolute or fixed $K^2$ values, which leads to potential uncertainties in subsequent calculations. Fortunately, the calculated evidence has shown that the mobility and statistically dynamics of the dye linker lead to a $K^2$ value of approximately 23 in almost all biological formats. The refractive index n of the medium has a value of 1.4 for biomolecules in aqueous solution. $Q_D$ is the quantum yield (QY) of the donor in the absence of the acceptor and J(λ) is the overlap integral, which represents the degree of spectral overlap between donor emission and the acceptor absorption. The values for J and $R_0$ increase with higher acceptor extinction coefficient and greater overlap between the donor emission spectrum and the acceptor spectrum. Whether FRET will be effective at a specific distance r can be estimated by the "rule of thumb" $R_0$ ± 50% $R_0$ for the upper and lower limits of the Förster distance. The efficiency of the energy transfer can be determined from either steady – state as in equation 2 or time resolved as in equation 3 measurements.

$$\mathrm{E} = 1 - \frac{F_{DA}}{F_D} \ldots\ldots\ldots\ldots\ldots\ldots\ldots\ldots\ldots\ldots\ldots\ldots. \text{(equation 2)}$$

$$\mathrm{E} = 1 - \frac{\tau_{DA}}{\tau_D} \ldots\ldots\ldots\ldots\ldots\ldots\ldots\ldots\ldots\ldots\ldots\ldots. \text{(equation 3)}$$

F is the relative donor fluorescence intensity in the absence $F_D$ and the presence $F_{DA}$ of the acceptor, and τ is the fluorescent lifetime of the donor in the absence D and presence DA of the acceptor [11].

### 2.5.2 Interactions Involved in Nonradiative Energy Transfer

Energy transfer can result from different interaction mechanisms. The exchange may be Coulombic and due to intermolecular orbital overlap as in Figure 8. The Coulombic interactions consist of long-range dipole-dipole interactions (Förster`s mechanism) and short-range multipolar interactions. The interactions due to intermolecular orbital overlap, which include electron exchange (Dexter mechanism) and charge resonance interactions, are of course only short range.

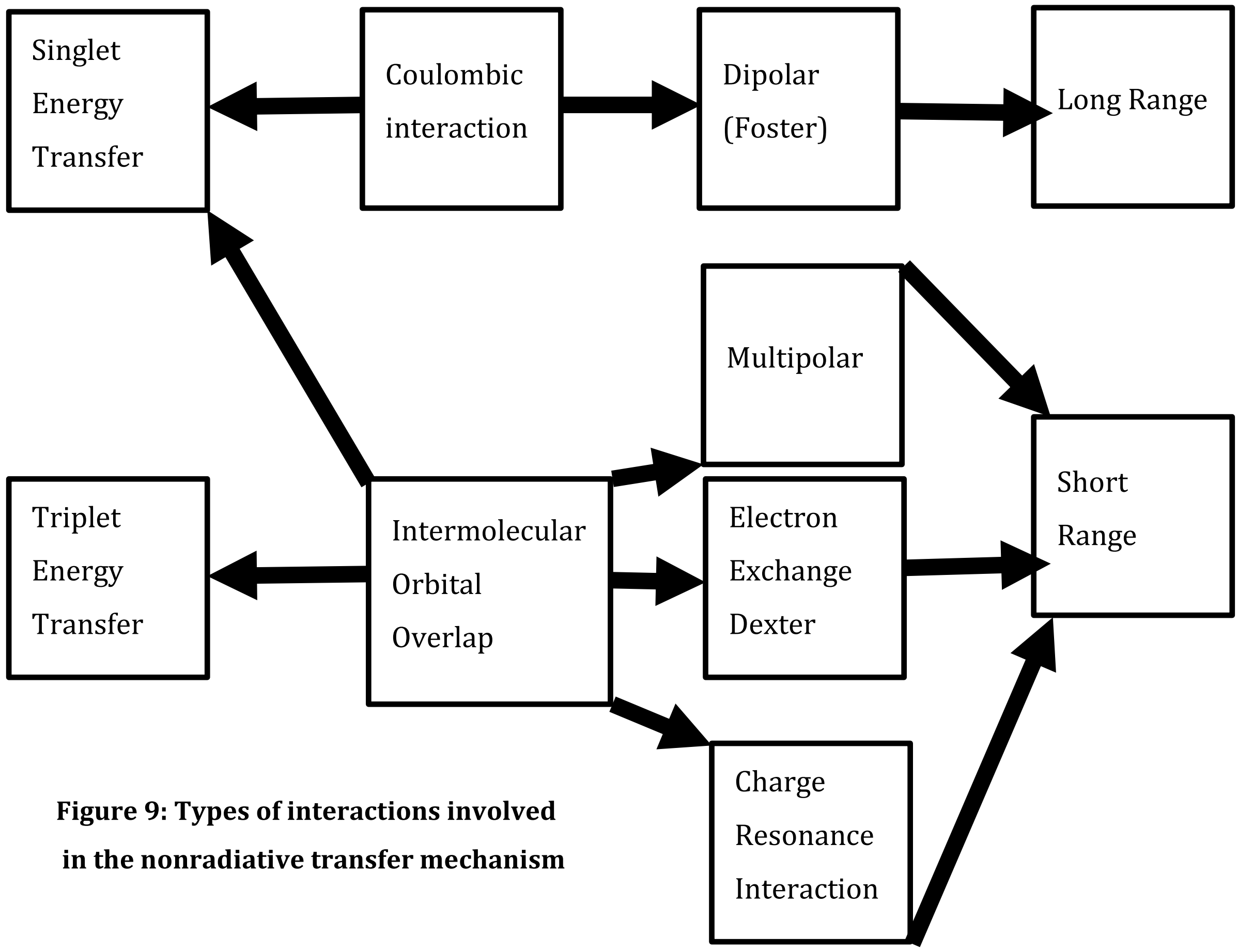

**Figure 9: Types of interactions involved in the nonradiative transfer mechanism**

# 2.6 Photoluminescence

Photoluminescence is light emission from any form of matter after absorption of photons (electromagnetic radiation). It is one of the many types of luminescence (light emission) and initiated by photoexcitation (i.e., photons that excite electrons to a higher energy level in an atom). After excitation, many different relaxation processes typically occur in which other photons radiate. The period between absorption and emission may vary ranging from short femtosecond-regime for discharge involving free carrier plasma in inorganic semiconductors up to milliseconds for phosphorescent processes in molecular systems, and under exceptional circumstances delay of release may even span to minutes or hours [12].

Photoluminescence spectroscopy is a contactless, non-destructive method of probing the electronic structure of materials. Light directed onto a sample, where it is absorbed and imparts excess energy into content in a process called photo-excitation. One way the specimen can dissipate this excess energy is through the emission of light, or luminescence [12].

### 2.6.1 NIR Fluorescence of SWCNTs for Optical Sensing

SWCNT are widely used in a variety of biomedical imaging modalities, including optical imaging, magnetic resonance imaging, and nuclear imaging. The magnetic and nuclear imaging rely mainly on the external labels or impurities in the CNT samples for imaging contrast. The focus here is the utilization of the inherent physical properties of the SWCNTs for optical imaging in biological systems [1].

The intrinsic optical properties of SWCNTs offer unique opportunities for optical biosensing. First, semi-conducting SWCNTs fluoresce at NIR wavelengths (900-1600). This wavelength range is within the "tissue transparency window," in which the optical absorption of biological tissues is minimal, bounded by haemoglobin and water absorption. SWCNT-based optical biosensors operated at these NIR wavelengths could outperform conventional optical biosensors operated at visible wavelengths because blood cells and tissues are strongly absorbing and auto fluorescent at visible wavelengths. Second, the fluorescence of SWCNTs does not photobleach, and the intensity of the of the fluorescence does not fluctuate. Thus, SWCNT-based sensors could persist indefinitely, ideal for long-term sensing. In comparison, most conventional fluorophores emit fluorescence at visible wavelengths and rapidly photobleach or fades, limiting their applications [1].

.

## 2.7 Ultraviolet/Visible/Near Infrared (UV/VIS/NIR) Spectroscopy

UV/VIS/NIR spectroscopy is the most accepted systematic approach because of the versatility and ability to detect almost every molecule. This spectroscopy method uses UV/VIS/NIR electromagnetic radiation, the light is passed through a sample and the transmittance of light by the sample is measured. The absorbance can be calculated from the transmittance (T). An absorbance spectrum that is obtained shows the absorbance of a compound at different wavelengths. The amount of absorbance at any wavelength is due to the chemical structure of the molecule [13].

UV/VIS/NIR spectroscopy can be used in a qualitative manner, to identify functional groups or confirm the identity of a compound by matching the absorbance spectrum. It can also be used in a qualitative fashion, as the concentration of the analyte is related to the absorbance using Beer`s law, $A = \varepsilon bC$ where $\varepsilon$ is the molar attenuation coefficient, b is path length, and C is concentration. [13]

UV/VIS/NIR spectroscopy is used to quantify the amount of DNA or protein in a sample for water analysis, and as a detector for many types of chromatography. Kinetics of chemical reactions are also measured with UV/VIS/NIR spectroscopy by taking repeated measurements overtime [13].

### 2.7.1 Principles

A spectrophotometer is a device for measuring the reflection or absorbance characteristics of a sample. The design mechanism necessitates that the wavelength of radiation to be studied must be a narrow "window". Table 2 illustrates the predetermined electromagnetic radiation wavelengths for the ultraviolet, visible and near infrared radiation.

**Table 2: Electromagnetic radiation wavelengths**

| | |
|---|---|
| Ultraviolet Radiation | 300 nm to 400 nm |
| Visible Radiation | 400 nm to 765 nm |
| Near infrared Radiation | 765 nm to 3200 nm |

## 2.8 Dye – (Astraphloxin)

Astraphloxin molecule is a polymethine dye with two cyanine groups linked by short polymethine chain as shown in Figure 10. Astraphloxin (pink powder) is a crystalline salt. In solutions, astraphloxin molecule has various conformations [14], and fluorescence lifetimes are strongly dependent on rigidity of used solvent [15]. In non-rigid solvents, the lifetimes of astraphloxin fluorescence is 30-600ps, while in rigid matrixes – up to 10ns [15].

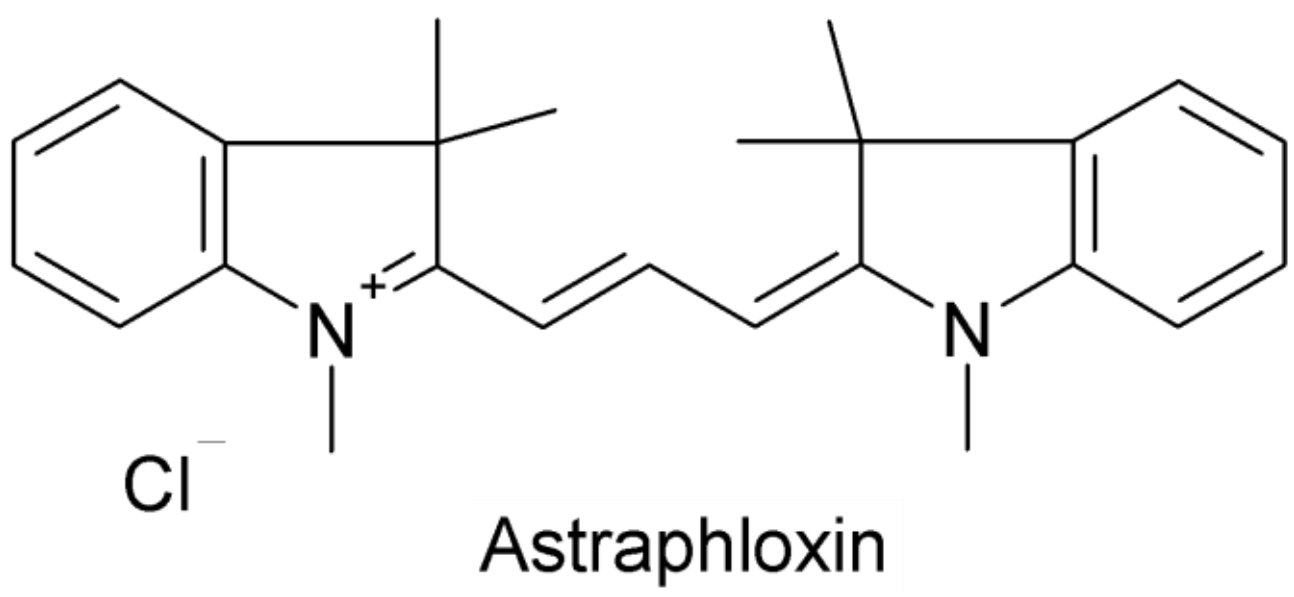

**Figure 10 – Molecular structure of astraphloxin [2].**

Essentially for this project, it was established that the astraphloxin molecule strongly interacts with SWCNTs and forms a single layer - shell around the nanotubes [2,16]. Figure 11 below can be used to explain the phenomenon of energy transfer in the mixture of the astraphloxin molecule and the CNT - surfactant.

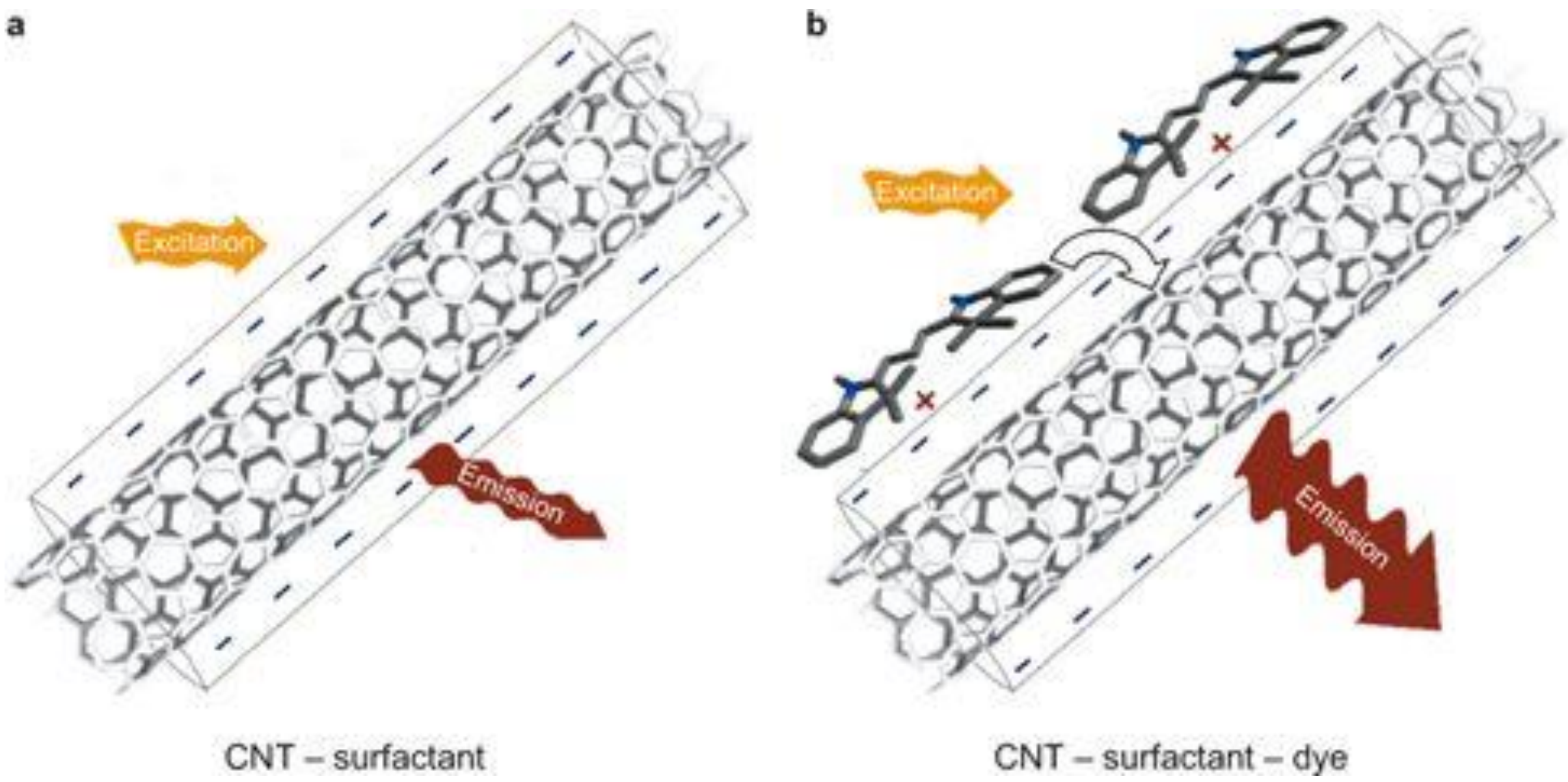

**Figure 11 – (a) The addition of a surfactant to the CNT helps in forming a micelle around it. (b) The mixture of CNT-surfactant and dye results in Coulomb attraction between the positively charged dye molecule of astraphloxin and CNT micelle [2].**

CNT-surfactant systems (Figure 11a) have very low photoluminescence. In the mixture of CNT with astraphloxin strong amplification of photoluminescence (up to 6 times) was observed from the nanotube levels in NIR range (Figure 11b). Thus, efficient energy transfer occurs from the astraphloxin to the tubes [2,16]. Other types of polymethine dyes were explored in study of energy transfer from the dye to CNT, and one of the most interesting results were obtained for the dioxaborine-cyannine dye (DOB-719) [17,18]. CNT with DOB-719 have efficient energy transfer photoluminescence with amplification more than an order of magnitude. Therefore, study of new modifications of cyanine dyes for functionalization of CNTs could improve the amplification of the nanotube emission further.

## 2.9 Dye - (Bis-astraphloxin)

Molecular structure of bis-astraphloxin is shown in Figure 12. This molecule consist of two astraphloxin units joined by a short linker (Figure 12). Our expectation is that the bis-astraphloxin could form a double layer – shell around the CNT resulting in enhanced photoluminescence (PL). This would improve the prospects of dye-CNT applications in imaging and the nanotube detection.

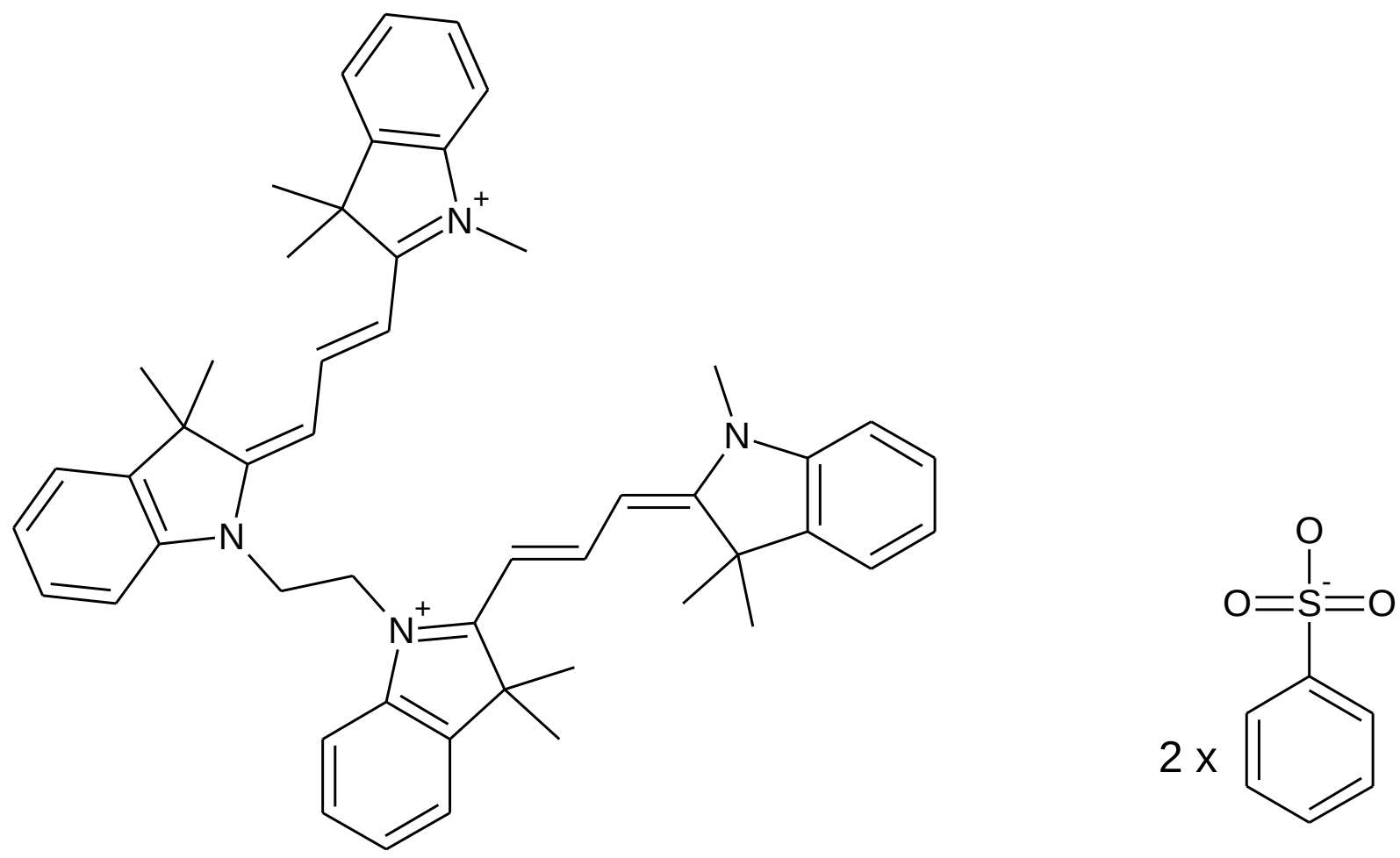

**Figure 12 – Molecular structure of bis-astraphloxin formed as double molecule of astraphloxin, the negatively charged counter-ion on the right-hand side is added to stabilize positively charged bis-astraphloxin.**

# 2.10 Surfactant – (SDBS)

One of the contributing factors that affect photoluminescence amplification is the type of ionic surfactant used. Therefore, SDBS as an essential surfactant is used in the experiment because of its high efficiency and prolonged stability in dispersing SWCNTs and thus yielding stable de bundled nanotubes in water [2].

# 3 Procedure

The methodology involved 2 main phases of the experimental setup, which are bulleted below and will be discussed but before discussing the setups, a step by step process of a sample preparation will be examined.

- Absorption characterization experimental setup
- Photoluminescence (PL) characterization experimental setup

## 3.1 Sample Preparation

In this experimental research accurate measurements of prepared samples were very crucial. The precision balance used as the primary instrument for weighting both the powder of SWCNT (CG100), SDBS, and dyes. This experimental approach employed a simple technique for preparing SWCNT-surfactant blend, whereby a 1.2 mg of SWCNT power dispersed in 20 mL of deionized water in the presence of 6.5 mg of anionic surfactant sodium dodecylbenzene sulfonate (SDBS). Initially, the above dispersions were subjected to ultrasonication using a NanoRuptor processor for one at 21 kHz, and 200 W. After these non-fluorescent SWCNTs gets sonicated, single SWCNTs suspended, and sonication separates SWCNTs, which becomes fluorescent. Furthermore, the dispersions were subjected to ultracentrifugation for 1 hour at 17$^{o}$C at 45,000 rpm. The process was done to separate larger particles of SWCNTs from the smaller particles. It also removes the aggregate phase and obtains de-bundled supernatant solutions of SWCNTs.

The following organic dye concentrations were calculated, measured and studied: 0.001, 0.004 and 0.008 mg/mL. But first and foremost, the precision balance was used to measure 0.77 mg of bis-astraphloxin to prepare the stock solution of the dye (bis-astraphloxin). The aim was to achieve a stock concentration of 0.1 g/L. Therefore, the simple calculation below was adopted to determine the volume of deionized water for the stock solution.

$$C = \frac{m}{V} \ldots\ldots\ldots\ldots\ldots\ldots\ldots\ldots\ldots\ldots\ldots\ldots. \text{(equation 4)}$$

In the above equation, $C$ is the stock concentration, $m$ is the mass of the dye (bis-astraphloxin), $V$ is the volume of deionized water

$$V = \frac{m}{C} = \frac{0.77\ mg}{0.1\ mg/ml} = 7.7ml$$

The prepared solution was then transferred into a sample holder bottle using a pipet. The container was placed in a shaker to fasten the dissolution of the bis-astraphloxin.

### 3.1.1 Sample Cell

The sample cell is a container where the mixtures are kept for experimental measurements. Energy transfer takes place in the sample cell when radiated by visible range electromagnetic radiation. When the donor (dye) in the sample cell absorbs light, electrons are advanced to excited energy states and transfer its energy from the high energy state by nonradiative process to another molecule (acceptor, in our case in SWCNT) and the acceptor emits light with photons of lower energy. In this experimental research, five main samples that were studied are shown in Figure 1 and Figure 2 and are also listed below:

1. Bis–astraphloxin solution
2. SWCNT + surfactant dispersion
3. Mixtures of SWCNT – surfactant – dye (bis–astraphloxin) at different concentrations.
4. Astraphloxin solution
5. Mixtures of SWCNT – surfactant – astraphloxin

### 3.1.2 SWCNT Sample Preparation

The preparation of SWCNT dispersion for the measurements is made in two ways. First sample is non-diluted (initial) SWCNT dispersion, which was used for absorption spectra measurements only: 3 mL non-diluted or initial SWCNT solution transferred into a sample cell holder 1. Second sample of diluted SWCNT dispersion is made to ensure a measurements of absorption and PL spectra for the mixtures and neat SWCNT has the same concentration of the nanotubes: diluted SWCNT dispersion is prepared by 20% of initial SWCNT and 80% deionized water also pipetted into sample cell 2

### 3.1.3 A mixture of Initial SWCNT with Organic Dye

The basic idea in this section was to demonstrate the process involved in calculating the different concentration contents as in Table 3 and to give insights to how the different concentration of the mixtures for neat SWCNT with organic dye was measured to give a total volume of 3mL. Therefore, equation (5) below was utilized for calculations and in the equation, $C_1$ is the stock concentration given as 0.1 g/L. $C_2$ is the studied organic dye concentrations given in g/L and $V_2$ is the total volume which is constant at 3 mL. To find the volume $V_1$ of the organic dye taken from stock solution ($C_1 = 0.1$ g/L), Equation 5 was used.

$$V_1 = \frac{C_2 V_2}{C_1} \ldots\ldots\ldots\ldots\ldots\ldots\ldots\ldots\ldots\ldots\ldots\ldots\ldots. \text{(Equation 5)}$$

Three different mixtures of dye (bis-astraphloxin) concentrations (0.001, 0.004 & 0.008 g/L,) were prepared for the new system as in Table 3 and four different mixtures of dye (astraphloxin) concentrations (0.0005, 0.001, 0.002 & 0.004 g/L) were prepared for the reference system.

**Table 3: Different sample mixtures and volume measurements for the SWNT-surfactant-dye**

| CNT (20%) | **Dye (0.0005 g/L)** | $H_20$ | Total Volume |
|---|---|---|---|
| 0.6 mL | **0.015 mL** | 2.385 mL | 3 mL |

| CNT (20%) | **Dye (0.001 g/L)** | $H_20$ | Total Volume |
|---|---|---|---|
| 0.6 mL | **0.03 mL** | 2.37 mL | 3 mL |

| CNT (20%) | **Dye (0.002 g/L)** | $H_20$ | Total Volume |
|---|---|---|---|
| 0.6 mL | **0.06 mL** | 2.37 mL | 3 mL |

| CNT (20%) | **Dye (0.004 g/L)** | $H_20$ | Total Volume |
|---|---|---|---|
| 0.6 mL | **0.12 mL** | 2.28 mL | 3 mL |

| CNT (20%) | **Dye (0.008 g/L)** | $H_20$ | Total Volume |
|---|---|---|---|
| 0.6 mL | **0.24 mL** | 2.16 mL | 3 mL |

# 3.2 Absorption Characterization Experimental Setup

The Lambda 1050 UV/VIS/NIR (Perkin Elmer) spectrometer shown in Figure 13 was used during the experiment to measure the absorption spectra in the UV/VIS/NIR.

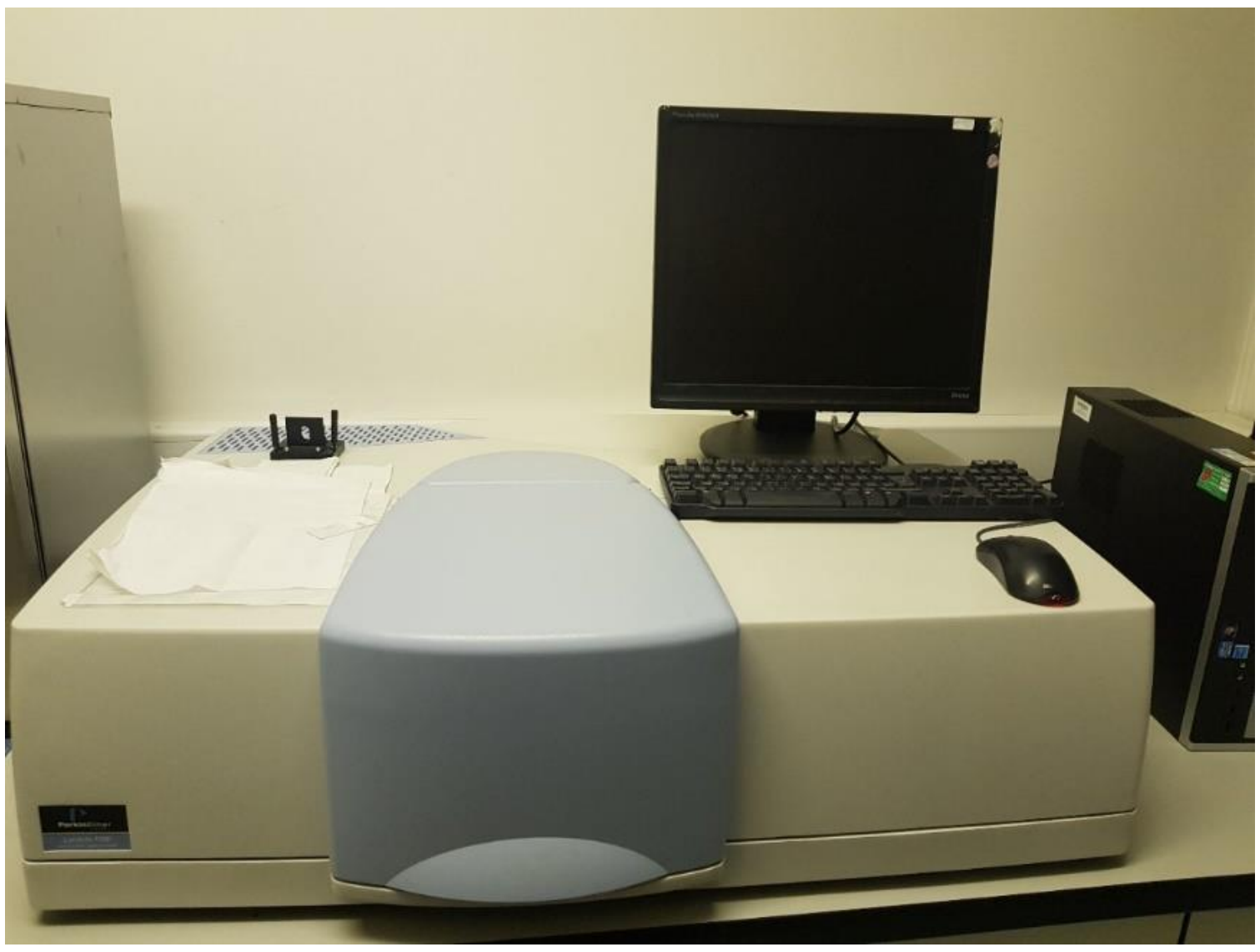

**Figure 13 – Photo of the Perkin Elmer 1050 Spectrometer with the sample compartment in the closed position and detector module installed.**

For scientific explanation, Figure 14 has revealed the schematic setup of the Perkin Elmer Lambda 1050 spectrophotometer. The Lambda 1050 UV/VIS/NIR spectrophotometer is a double slit spectrometer. Measurements made in a spectral range from 250 nm to 1400 nm, which is covered by two lamps. The location of the two lights is shown in Figure 14 above. A deuterium lamp $D_2$ for UV section and a tungsten lamp W for the visible and NIR spectral regions. For absorption spectra, measurements in the visible and NIR ranges the tungsten lamp was used. The tungsten lamp produces, like the incandescent light bulbs, a continuous spectrum from the near UV throughout the rest of the spectral range. The light was imaged through a slit assembly towards a set of filters and two consecutive monochromators which use holographic gratings. The monochromators filter a fraction with a small bandwidth. The light beam continues through a common beam mask (CBM), which can set the spot size. The CBM depolarizer follows, which can be included in the measurements if required. Thereafter, the light beam is separated into the reference beam and sample beam by a chopper, consisting of a rotating disc with a mirror and a window segment directing the beam in either direction. The two beams are then directed through the sample compartment of the spectrometer, where the samples are placed in either path. The detector module installed at the end of the schematic diagram for detection [13].

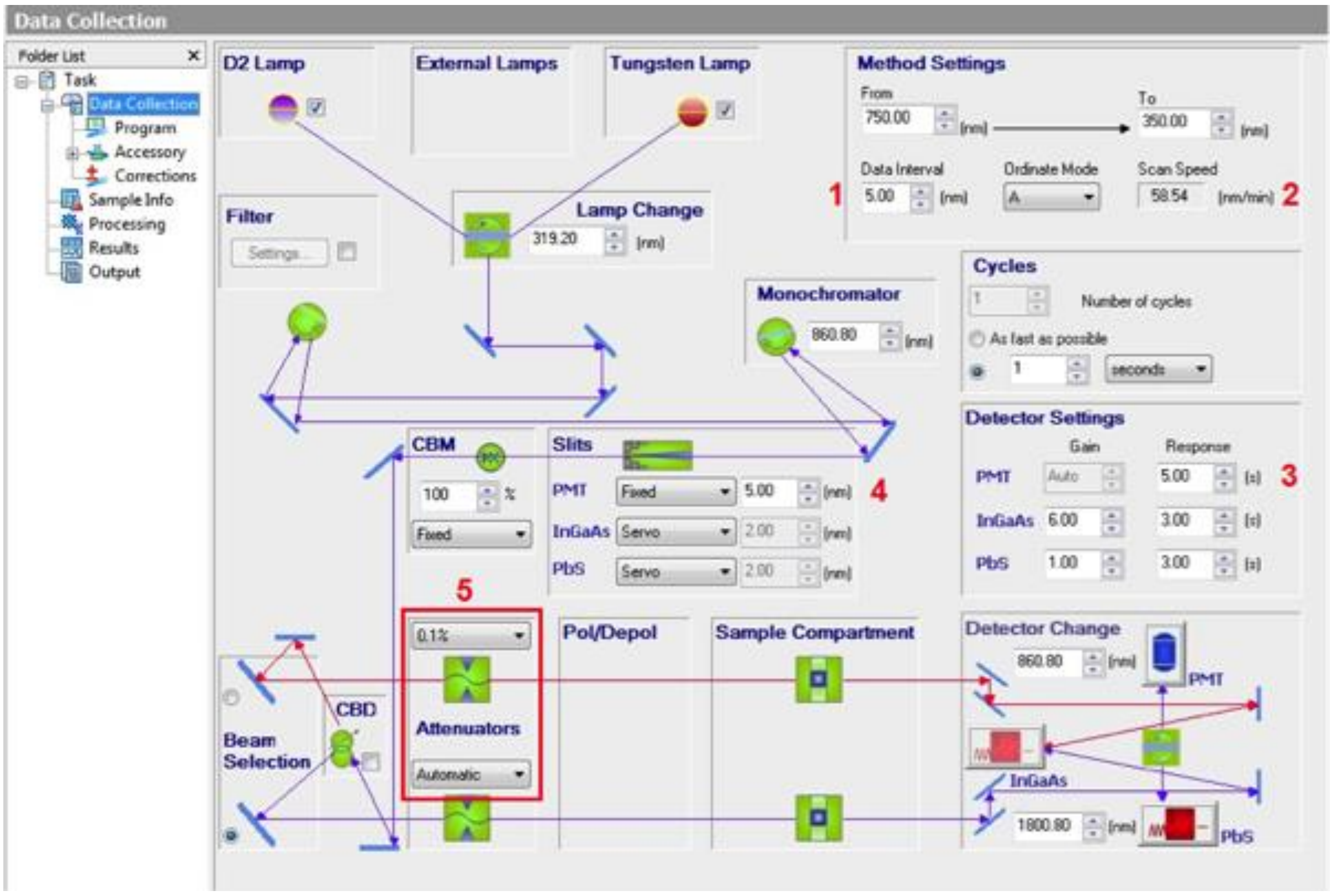

**Figure 14 – Path of light through the schematic setup of the Perkin Elmer Spectrometer.**

## 3.3 Photoluminescence (PL) Characterization Experimental Setup

The fluorescence emission spectra at various excitation wavelengths were recorded using Horiba NanoLog excitation – emission spectrofluorometer equipped with nitrogen – cooled InGaAs array detector to generate PL excitation-emission maps as in Figure 15.

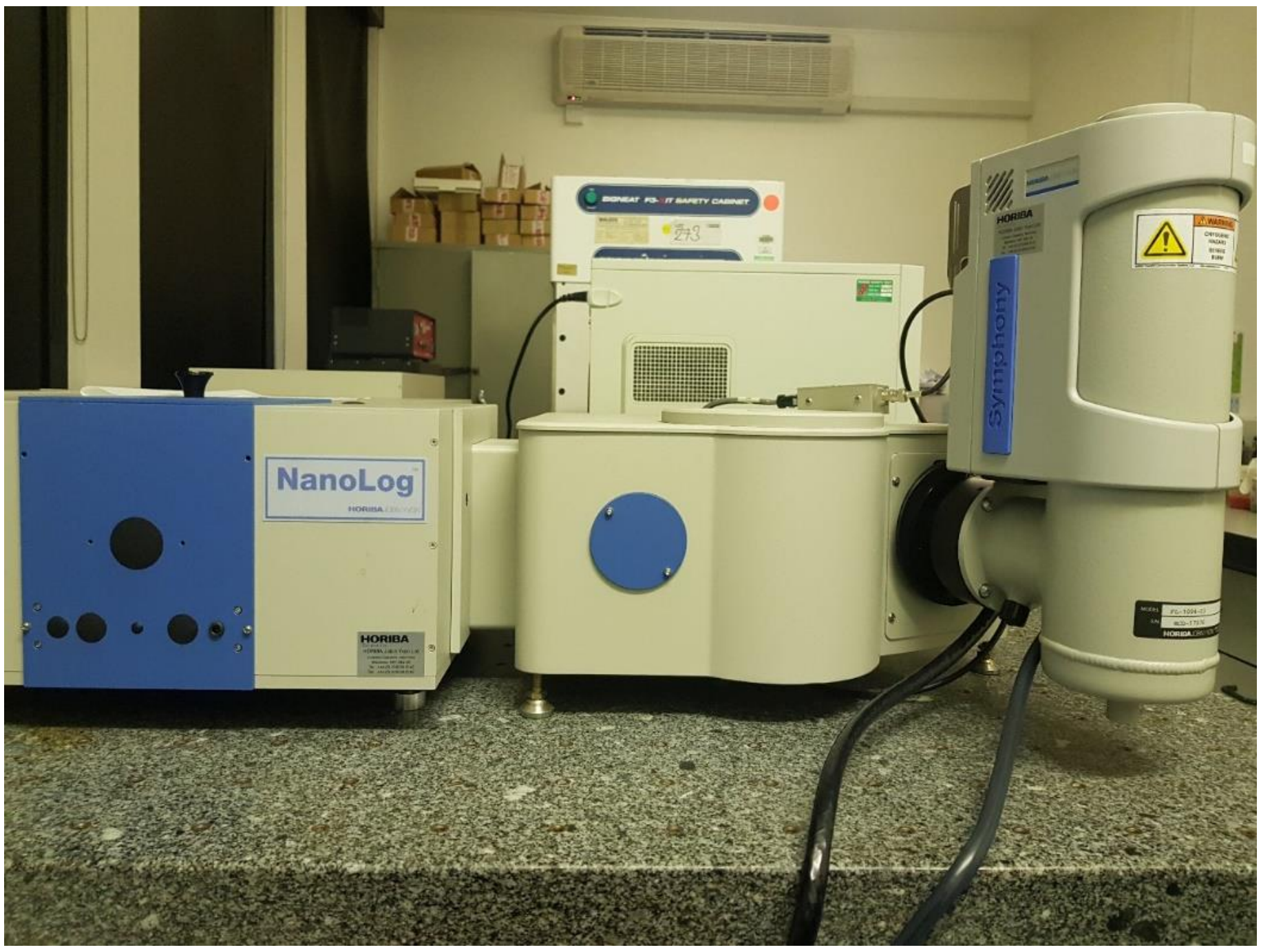

**Figure 15 – Fluorescence Spectrometer**

Figure 16 represents a simplified diagram of a spectrofluorometer for the purpose of comprehension.

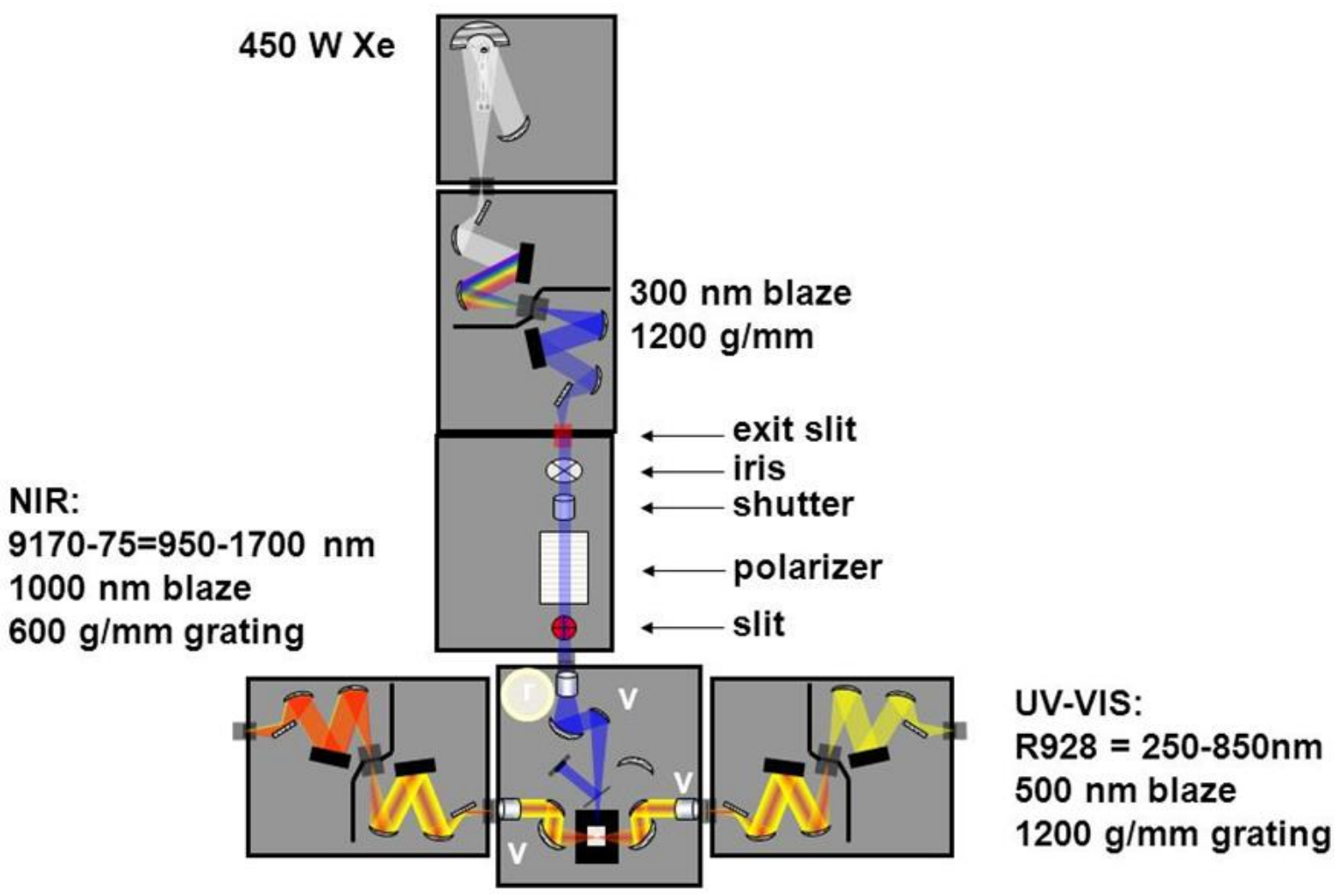

**Figure 16 – Fluorolog – 3 Spectrofluorometer**

Such system allows to record PLE maps, where excitation wavelength ( $\lambda_{EX}$ or EX wavelength) is plotted on the Y-axis, emission wavelength ( $\lambda_{EM}$ or EM wavelength) on the X-axis, and PL intensity represented as a contour map. The highest intensity is shown in red, whereas blue colour shows the lowest PL intensity.

The 450 W excitation is the source of radiation for producing photons. The beam of light that comes from the lamp is filtered by an excitation monochromator that allows only a single wavelength of light to reach the sample. The sample compartment module is a T- box, which provides efficient throughput with a choice of standard right – angle emission collection. The sample compartment module comes equipped with a silicon photodiode reference detector to monitor and compensate for variations in the source output. The emission from the sample is filtered by an emission monochromator that feeds the signal to a photomultiplier detector. By stepping either or both monochromators through a wavelength region and recording the variation in intensity as a function of wavelength, a spectrum gets produced [8]. The monochromator, sample – compartment unit and accessories are all connected to a controller called the SpectrAcq, which transfers information to and from the host computer.

# 4 Experiments

For absorption and photoluminescence characterization, the experiments were conducted by first measuring absorption spectra in the visible and NIR ranges for the samples and second by measuring the fluorescence emission spectra in the NIR or visible ranges for the initial CNT, diluted CNT (20% CNT and 80% water), mixtures of SWCNTs and dyes (bis-astraphloxin or astraphloxin) at various concentrations (Table 3).

## 4.1 Measuring Absorption

After sample preparation as can be seen in Figure 17. The absorption spectra in the visible and NIR ranges are measured by using the Lambda 1050 UV/VIS/NIR spectrometer and a host computer.

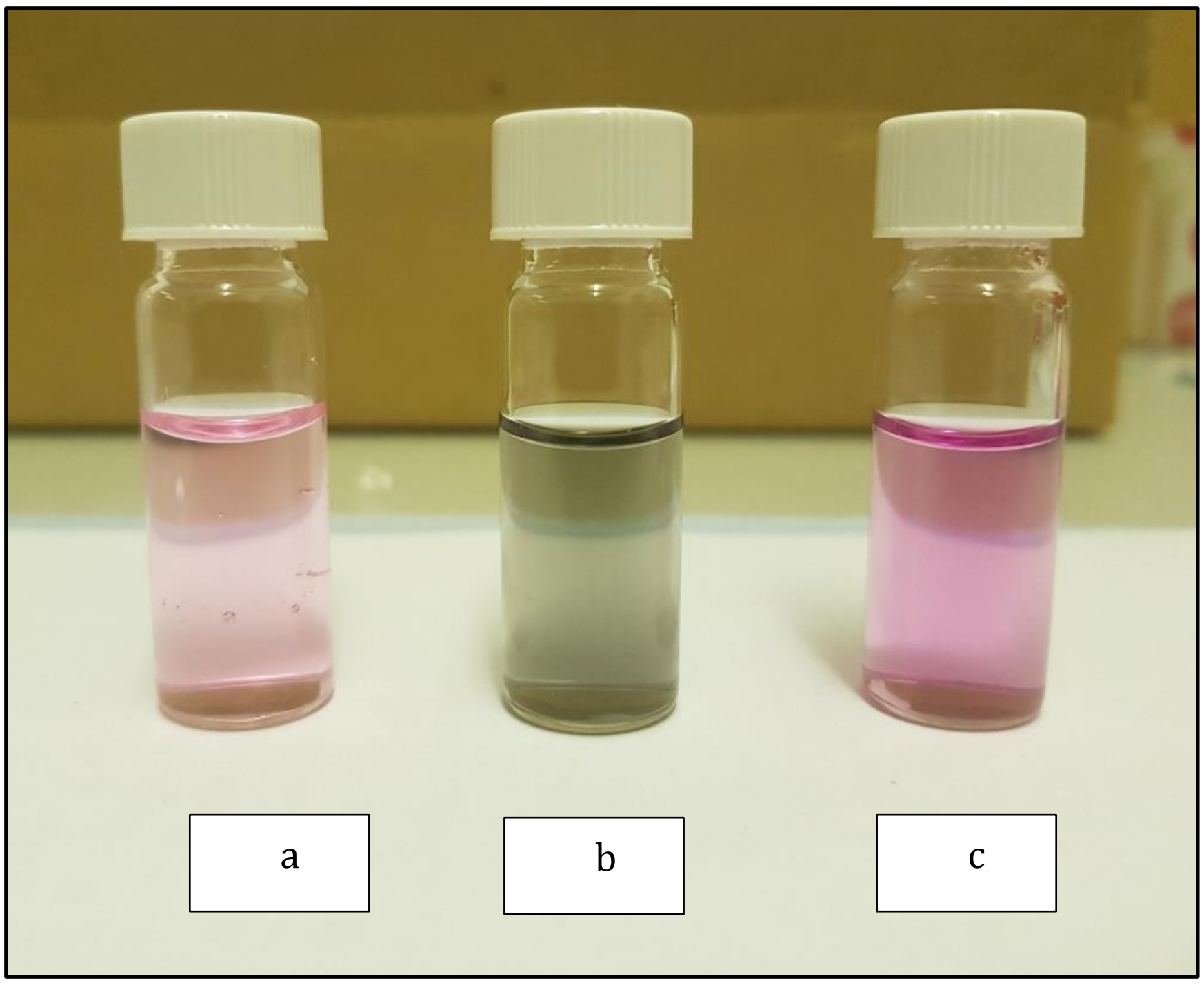

**Figure 17 – Prepared samples (a) Mixture of CNT + Bis-Astra 0.001 g/L, (b) CNT - surfactant (c) Mixture of CNT + Bis-Astra 0.004 g/L**

The process of setting the parameters for measuring the absorptions spectra comes after all the devices and accessories are connected to the host computer as shown in Figure 14. The procedures involved are listed below.

1. Calibration of the spectrometer to establish zero line (or 100% transmission)

2. Performing immediate measurements of an absorption spectrum
3. Performing delayed measurements of an absorption spectrum

### 4.1.1 Zero-line calibration of the spectrometer

Calibration of zero-line is an essential element for accurate measurements to be achieved. The step-by step procedure taken to calibrate the spectrometer without samples was straight forward and described below.

- Turned on the spectrometer and allowed the lamp to warm up for an appropriate time (about 15 to 20 minutes) for stabilization of the system.
- Double clicked PerkinElmer on the desktop of the host computer
- Clicked on Scan lambda to begin data collection and method settings set between 1400 nm (infrared) to 250 nm (UV).
- The data interval set to either 1 or 1.50 (steps) depending on the sample
- Clicked on sample information to set the number of samples
- Clicked start for zero line

### 4.1.2 Immediate Measurements of Absorption

Before measuring, each sample was freshly prepared and placed in a sample cell for instant measurements. The absorption spectra of each sample were collected by setting the parameters and placing the sample cell in the sample compartment and deionized water as a reference sample which was placed in the reference sample compartment. Both samples are cleaned of any fingerprints. The sample cell and the sample compartment are properly closed to prevent any ambient light. Collected an absorption spectrum by allowing the device to scan through different wavelengths and collected the absorbance. From the collected spectrum, the absorbance maxima were determined ($\lambda_{max}$). Repeated the collection of spectra to get estimated error in the ($\lambda_{max}$).

### 4.1.3 Delayed or Temporal Measurements of Absorption

The same technique of measuring samples in “immediate measurements” are applied to delayed measurements as well, except that the sample cells are aged over time and measured to investigate the behaviour of the systems and the energy transfer that takes place after few days to a week.

# 4.2 Results

## 4.2.1 Immediate Measurement of Dye and CNT

The absorption spectra of the dyes and CNTs as seen in Figure 18, were first taken to gain some essential information about each concentration.

a)

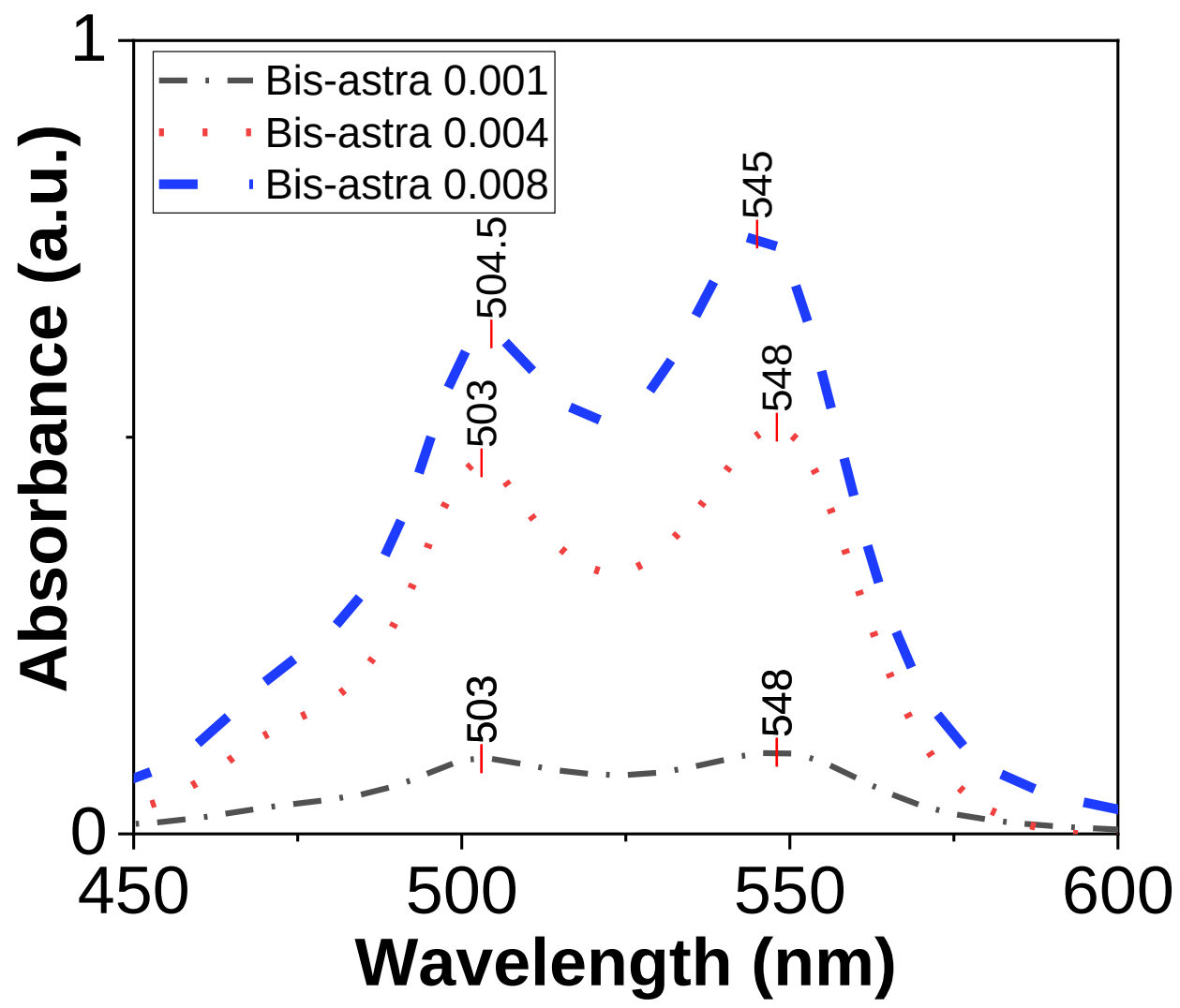

b)

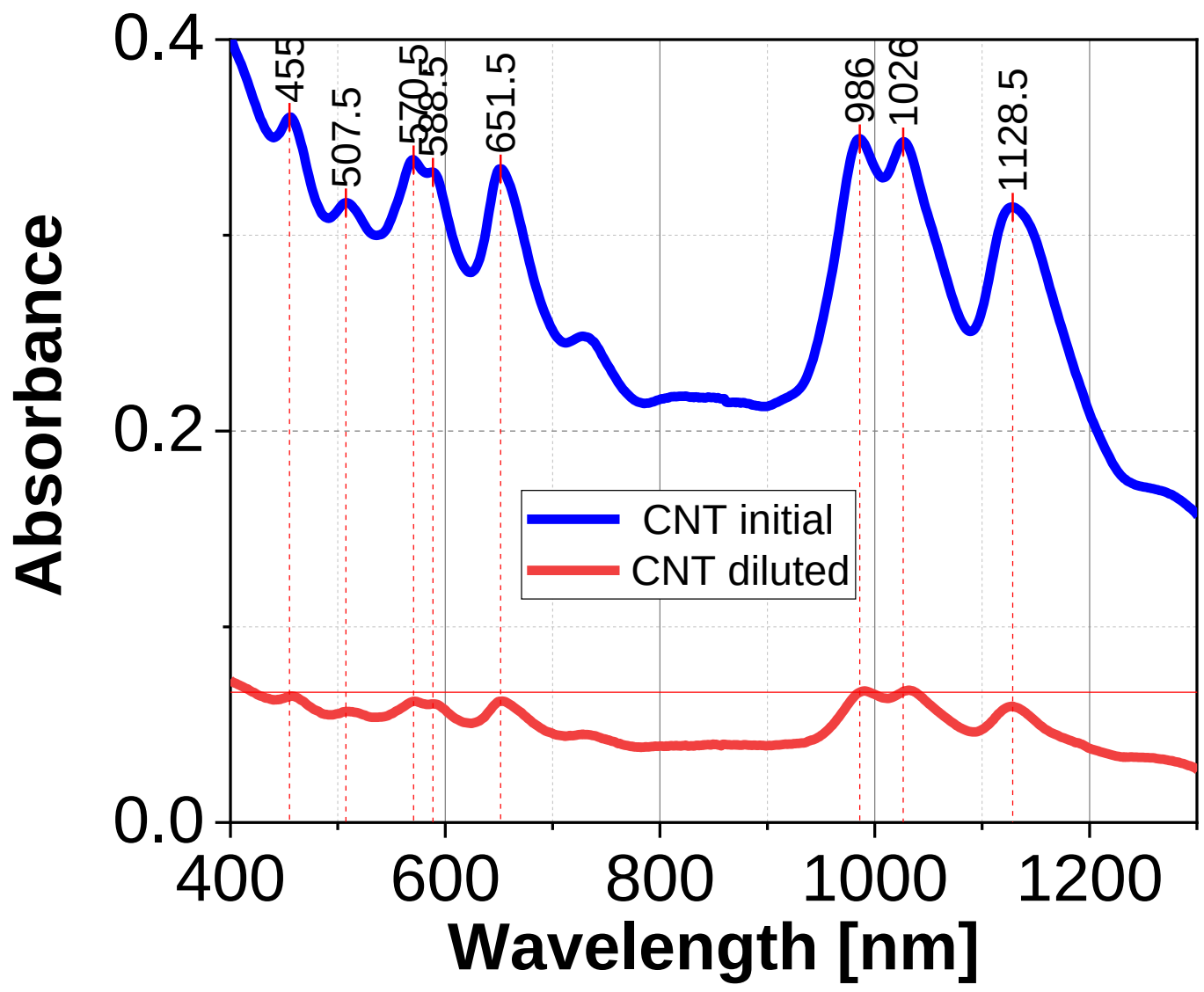

c)

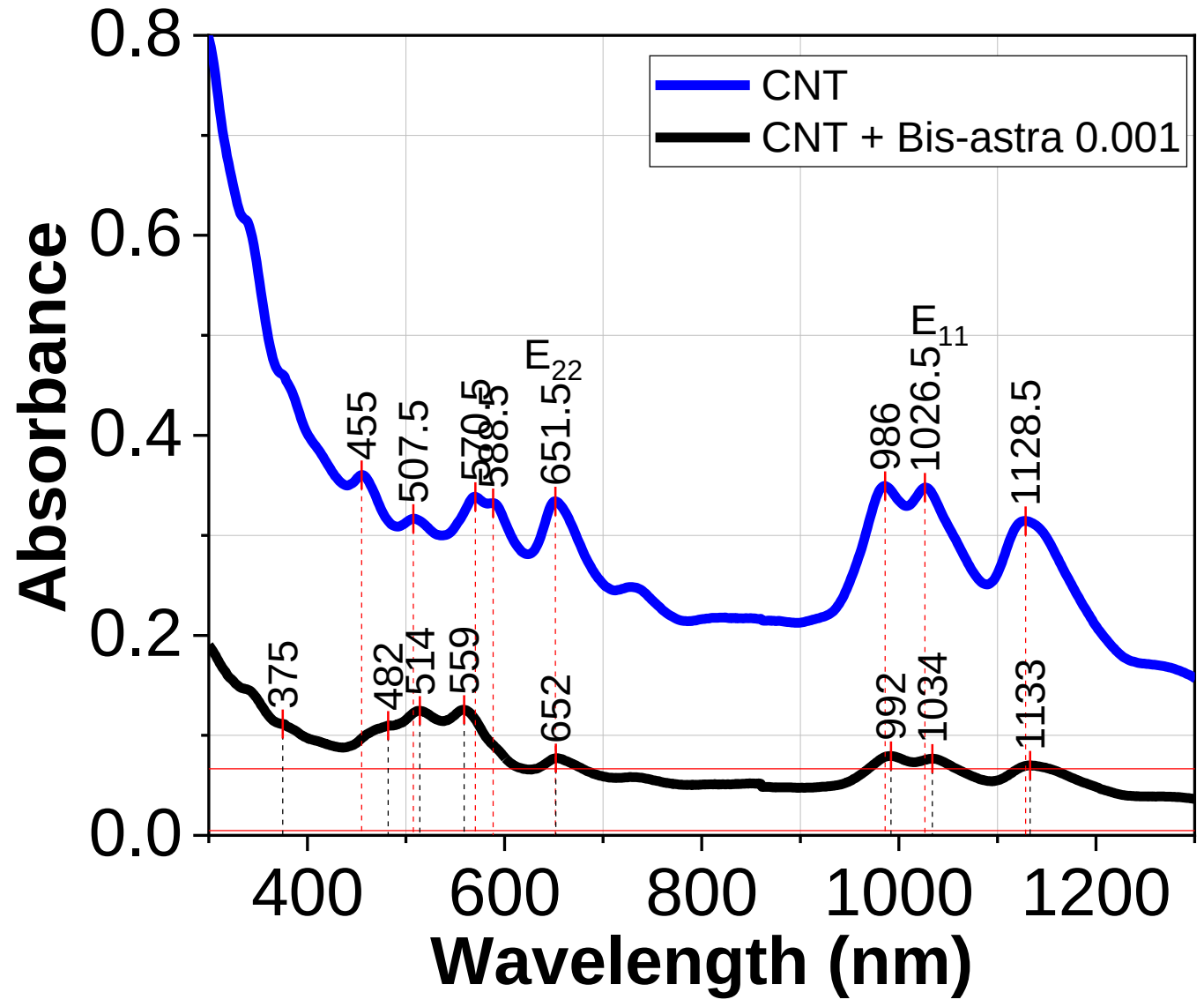

**Figure 18 – (a) Absorption spectra of water solutions of bis-astraphloxin (0.001 mg/mL, 0.004 mg/mL, 0.008 mg/mL). (b) Absorption spectra in the visible and NIR ranges of neat CNT dispersions: stock dispersion in blue and diluted 1:4 in red. (c) Mixture of CNT with bis-astraphloxin (0.001 mg/mL) in black and stock dispersion of CNT in blue. The CNTs dispersed in water with SDBS.**

### 4.2.2 Immediate Measurement of the Mixtures

The two absorption graphs in Figure 19, illustrated results for the freshly prepared diluted CNT, bis-astraphloxin, astraphoxin, mixtures of bis-astraphloxin + CNT and mixtures of astraphloxin + CNT.

a)

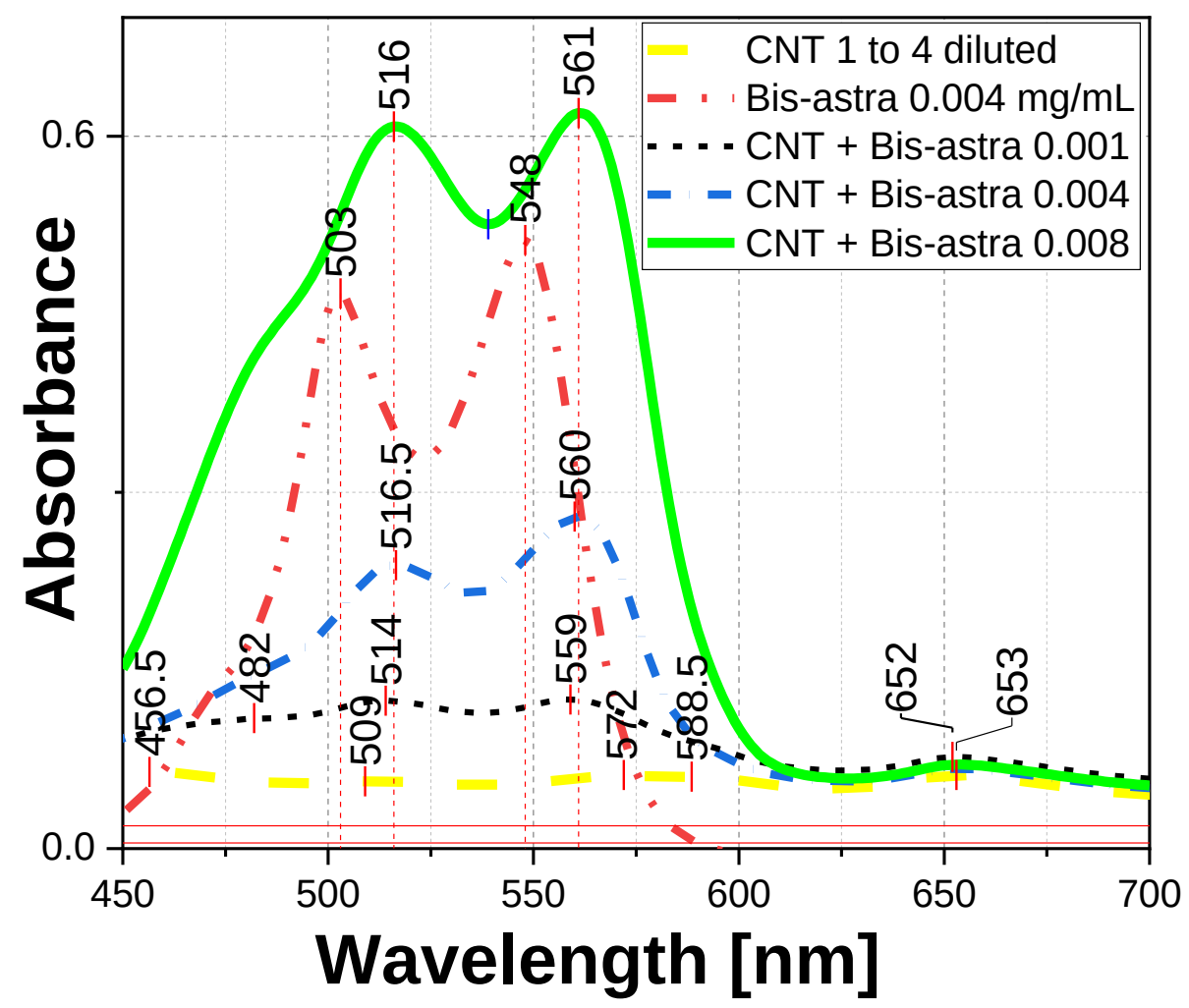

b)

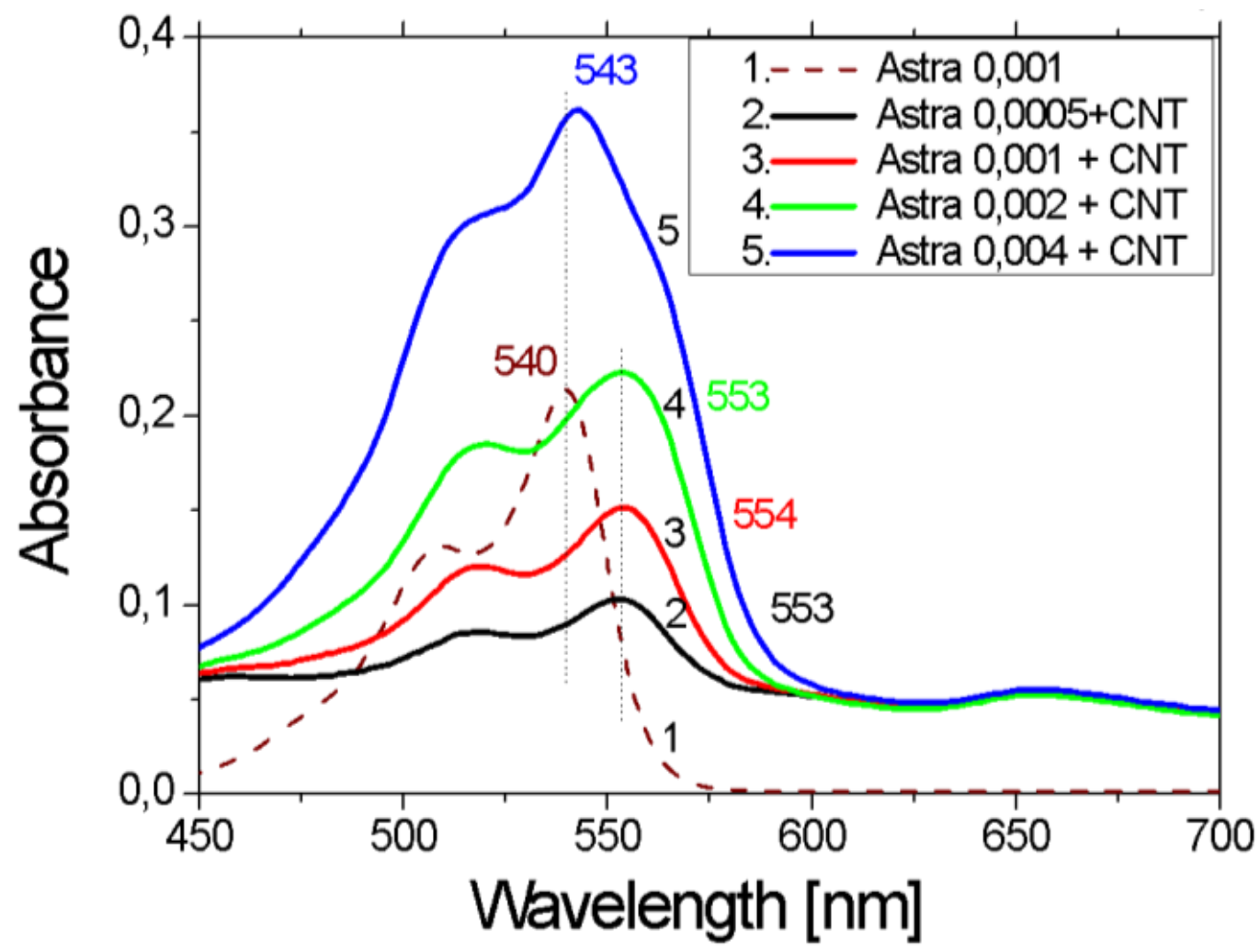

**Figure 19 – Absorption spectra of bis-astraphloxin, astraphloxin and SWCNTs dispersed with SDBS. (a) Absorption spectra of a mixture of 20 % CNT and 80% water (CNT 1 to 4 diluted) as reference, spectra of water solution of bis-astraphloxin (0.004 mg/mL), mixtures of CNTs with bis-astraphloxin (0.001 mg/mL, 0.004 mg/mL, 0.008 mg/mL, where the concentrations given here are the dye concentrations in the mixture). (b) Absorption spectra of water solution of astraphloxin 0.004 mg/mL, mixtures of CNTs with astraphoxin (0.0005 mg/mL, 0.001 mg/mL, 0.002 mg/mL, 0.004 mg/mL, here again the concentrations given are the dye concentrations in the mixture.**

### 4.2.3 Delayed/Temporal Measurement Results

Since the focus in this experimental research is to investigate energy transfer in the new system of CNT-surfactant-dye(bis-astraphloxin), that is expected to show an exceptional magnification of photoluminescence response via energy transfer and such attributes have many practical or industrial applications. The need to investigate energy transfer overtime is also very paramount and Figure 20 illustrates the absorption spectra of the samples after a few days to a week or more.

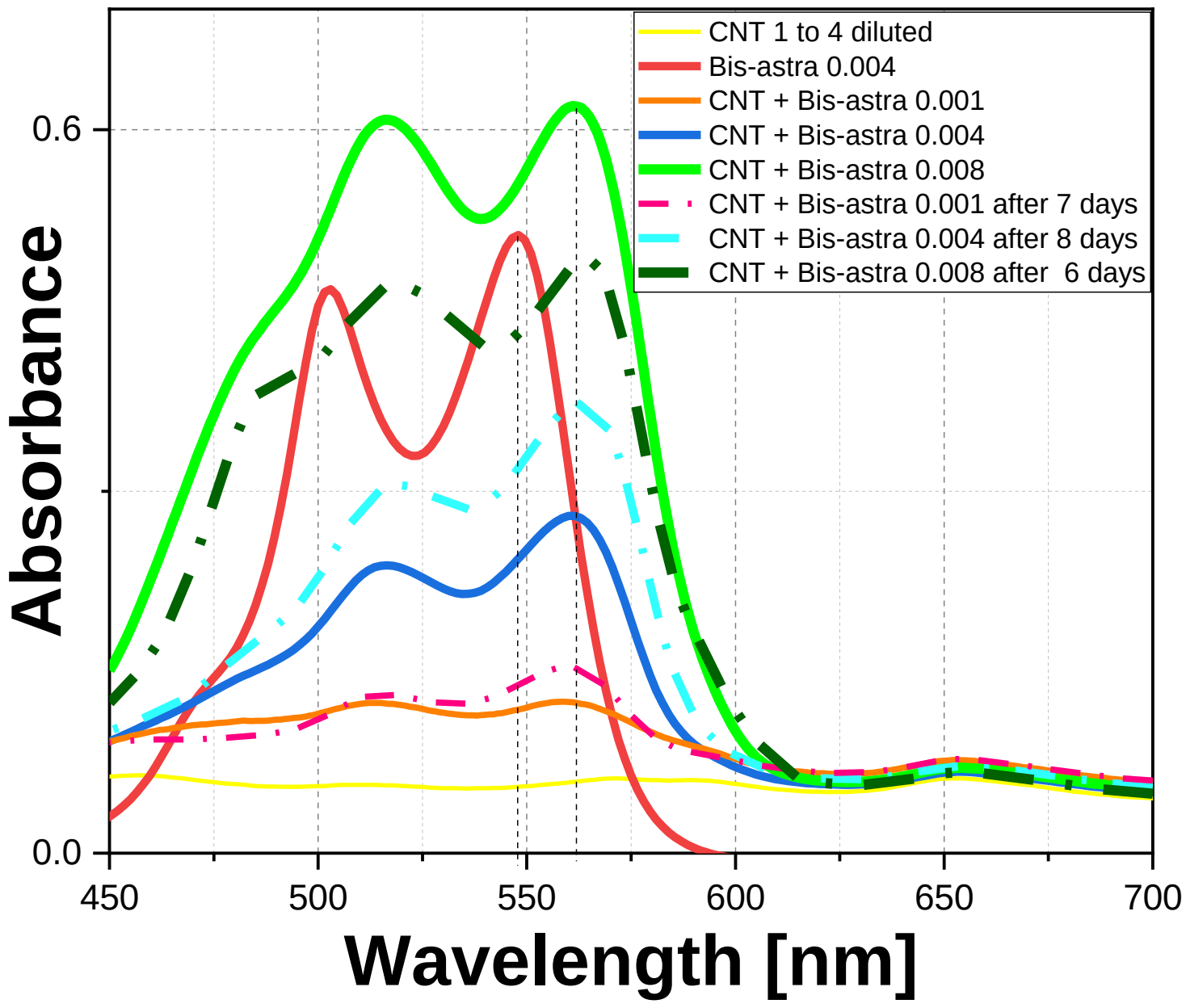

**Figure 20 –Absorption spectra of a mixture of 20 % CNT and 80% water (CNT 1 to 4 diluted) as reference, spectra of water solution of bis-astraphloxin (0.004 mg/mL), mixtures of CNTs with bis-astraphloxin (0.001 mg/mL, 0.004 mg/mL, 0.008 mg/mL, where the concentrations given here are the dye concentrations in the mixture). the mixtures of CNTs with bis-astraphloxin (0.001 mg/mL, 0.004 mg/mL, 0.008 mg/mL, where the concentrations are the dye concentrations in the mixture) aged over days to a week to investigate the energy transfer.**

## 4.3 Discussion/Analysis

The absorption spectra of bis-astraphloxin solutions as shown in Figure 18 (a), exhibit a maximum at $\lambda_M$ = 548 nm and a vibrational progression band at 503 nm for the two lower concentrations (0.001 mg/mL, 0.004 mg/mL). The absorption spectrum of bis-astraphloxin at higher concentration (0.008 mg/mL) has a blue-shifted $\lambda_M$ = 545 nm evidencing tendency for H-type aggregation of bis-astraphloxin molecules at high concentration. In should be noted that H-aggregate molecular alignment was observed for astraphloxin at much higher concentration [16]. This type of aggregation is ideally non fluorescent in nature. Normally such aggregation behaviour of fluorescent dye molecules strongly quenches their fluorescence, which has an adverse effect in designing display devices and sensors as this application are mainly based on the fluorescence properties. Therefore, it is very crucial for the applicability of these fluorescent dyes either to control aggregation or to get fluorescent signal from these aggregates. Aggregation of the dye molecules can be controlled by mixing these dyes with different materials [1, 16].

Figure 18 (b,c) illustrates experimental evidences for the interaction of bis-astraphloxin with CNT due to comparison of absorption spectra of neat CNT and mixture of CNT with the dye. The interaction is evidenced by the absorbance data in the field of transparency of the dye (600-1250nm): by analysing the excitonic peaks $E_{11}$ (900 nm – 1250 nm) and $E_{22}$ (550 nm -750 nm) in the CNT absorption spectra and the mixtures. It was observed that the high-energy $E_{22}$ peaks of the CNTs with its maximum at 652 nm, remains fixed in position. Whereas, the low energy $E_{11}$ bands peaked at 986 nm, 1026.5 nm and 1128.5 nm for neat CNT are red-shifted to 992 nm, 1034 nm and 1133 nm in the mixtures, respectively.

Experimental results for the immediate measurements displayed in Figure 19 (a) for CNT, bis-astraphloxin, and mixtures of CNT with bis-astraphloxin indicates the complexation of bis-astraphloxin with CNTs. The absorption spectra for the combinations of bis-astraphloxin with CNTs contained strongly red-shifted dye absorption peaks. The extent of the red-shift depends on the dye concentration. The strongest red shift was observed for dye concentration 0.008 mg/mL peaking at $\lambda_M = 561$ nm. The red-shifted absorption bands in the range of 555-600 nm referenced to J-like aggregates of bis-astraphloxin. We observe no peak of the monomers ($\lambda_M = 548$ nm) in the spectra of the mixtures. Thus, all molecules of bis-astraphloxin in the studied mixtures are interacting with CNT, and no monomers are present in the mixtures. We have compared the results for the mixtures with bis-astraphloxin and mixtures with astraphloxin. The mixture of CNT with astraphloxin molecule also illustrated the complexation of astraphloxin with CNTs as was observed previously [2,16]: the absorption spectra for the mixtures indicates red-shifted dye absorption peaks. However, the extent of the red-shift depended on the dye concentration. Importantly, the monomer molecules of astraphloxin at $\lambda_M = 540$ nm appear in the mixtures at concentration of 0.004 mg/mL. By observing the threshold in the appearance of the dye monomer, which falls in the range of 0.002-0.004 mg/mL, suggests that the astraphloxin covers the surface of the CNTs in a single layer. Considering that all bis-astraphloxin molecules interact with the nanotubes at much higher concentration, we assume that bis-astraphloxin is covering CNT with multi-layered shell: two or even more layers. This could be attributed to complex molecular structure of bis-astraphloxin molecules.

Temporal or delayed measurements of the absorption spectra of the neat dye (bis-astraphloxin or astraphloxin) and neat CNT dispersions aged over a week did not show any changes in the spectra (Figure 20). Besides, the mixtures of CNT-bis-astraphloxin did indicate some significant changes over time. The absorption of mixtures with low concentration (0.001 mg/mL) of the dye grows and is slightly red-shifted, when aged over a week. A mixture of CNT-bis-astraphloxin at 0.004 mg/mL aged over eight days manifested a considerable increase in the absorption intensity and the peaks are slightly red-shifted. Finally, the mixture at high concentration of the dye (0.008 mg/mL) when aged over six days behaves differently and instead decreases in absorption intensity. Such changes could be explained by improved complexation with time for the dye molecules at the concentrations equal and below 0.004 mg/mL. Oppositely, at higher concentrations (above 0.004 mg/mL) the process of dye complexation on the nanotubes could be interfered by creation of heavily covered CNT-dye systems, which start to form bundles. Such dynamics could be confirmed by PL measurements for aged samples of the mixtures.

# 4.4 Measuring Fluorescence Emission/ MAPS

As soon as samples were prepared as shown in Figure 17, the next procedure on measurements using the fluorescence spectrometer (Figures 15,16) was followed:

1. Fluorescence system calibration
2. Immediate measuring of samples in the VIS/NIR
3. Delayed/Temporal measurements of Samples in the VIS/NIR

## 4.4.1 Fluorescence Spectrometer System Calibration

The fluorescence spectrometer has two calibration checks that need to be done before running any measurements. This check helps to prevent errors in the measurements. The device is calibrated by conducting an excitation and emission calibration check. The excitation calibration check verifies the wavelength calibration of the excitation monochromator, using the reference photodiode located before the sample compartment. It is an excitation scan of the Xenon lamp output and should be a very initial check. The second calibration check is the emission check of the instrument, and it is directly affected by the calibration of the excitation monochromator. The emission calibration verifies the wavelength calibration of the emission monochromator with the emission photomultiplier tube.

## 4.4.2 Immediate measurement of Samples in VIS/NIR

The method of measuring in the visible and near infrared comes after all other system checks are conducted and indicating that the system is running correctly and no significant issues. The procedures adopted for measurements in both the visible and infrared are described below.

- Inserted a sample cell in the sample compartment and closed the lid of the compartment properly to avoid ambient light interfering with the specimen.
- Doubled clicked on the Icon on the desktop
- On the main FluorEssence toolbar, Selected the experiment menu button "**M".** This opens the main experiment window
- Selected the spectra button for the experiment type window to appear

- Specified by choosing either excitation or emission depending on what the measurement was. Then clicked the next button.
- Adjusted the default parameters for the monochromator depending on the measurement.
- Clicked the Run button "**Run**"

Table 4 and Table 5 below indicate the setting parameters for performed measurements in either the NIR or visible range.

**Table 4: Settings for PL measurements in the visible range**

| **Monochromator parameters for measurements in the visible range** | | | | |
|---|---|---|---|---|
| Monochromator (1200 grooves/mm) | Initial Wavelength | Final wavelength | Increment | Slits |
| Excitation | 400 nm | 700 nm | 5 nm | 2 nm |
| Emission | 450 nm | 800 nm | 5 nm | 2 nm |

**Table 5: Settings for PL measurements in the infrared range**

| **Monochromator parameters for measurements in the infrared range** | | | | |
|---|---|---|---|---|
| Monochromator (1200 grooves/mm) | Initial Wavelength | Final wavelength | Increment | Slits |
| Excitation | 300 nm | 800 nm | 5 nm | 10 nm |
| Emission | 900 nm | 1400 nm | 5 nm | 10 nm |

# 4.5 Results

## 4.5.1 Immediate Measurement of PLE Maps of Bis-astraphloxin and Astraphloxin

a)

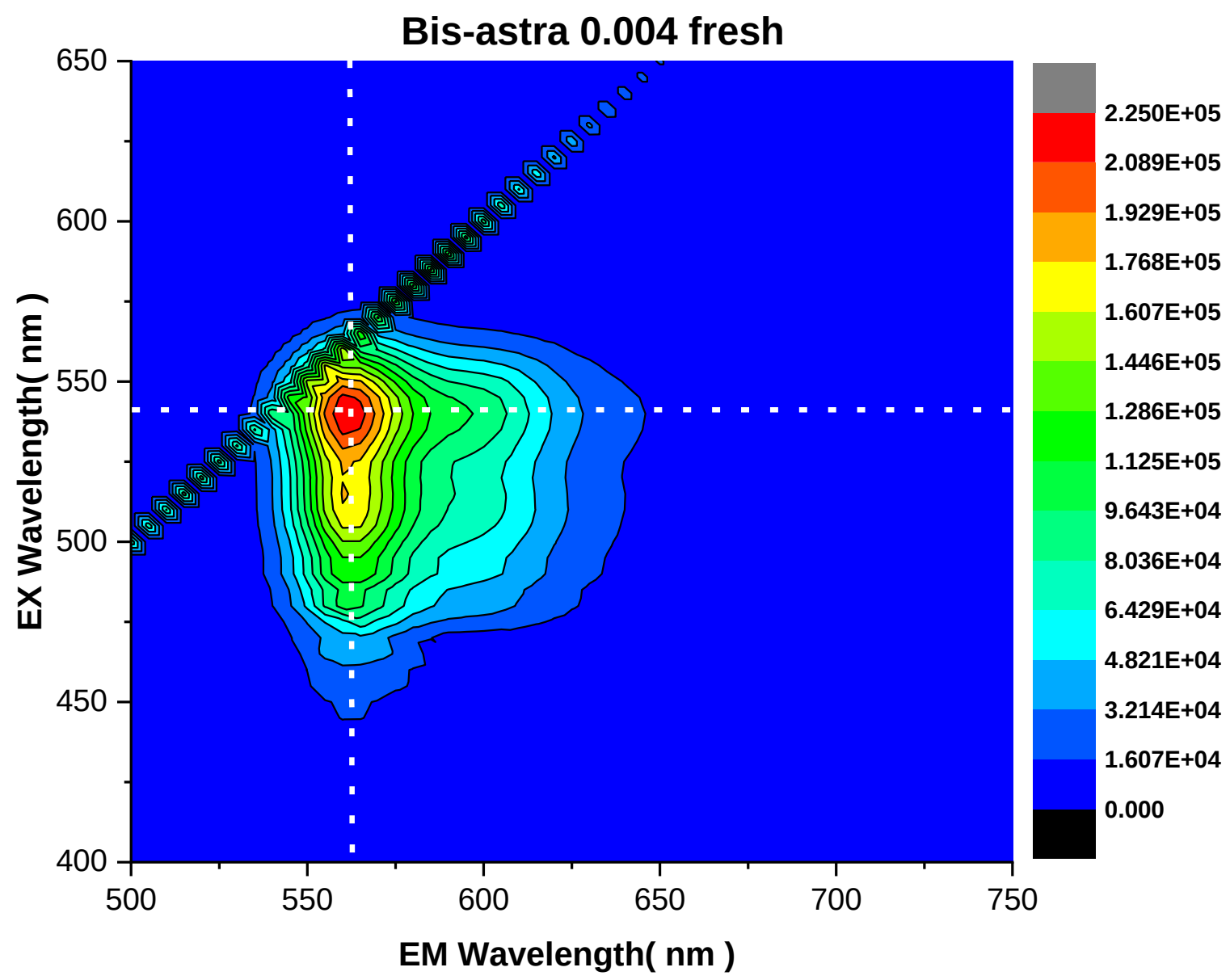

b)

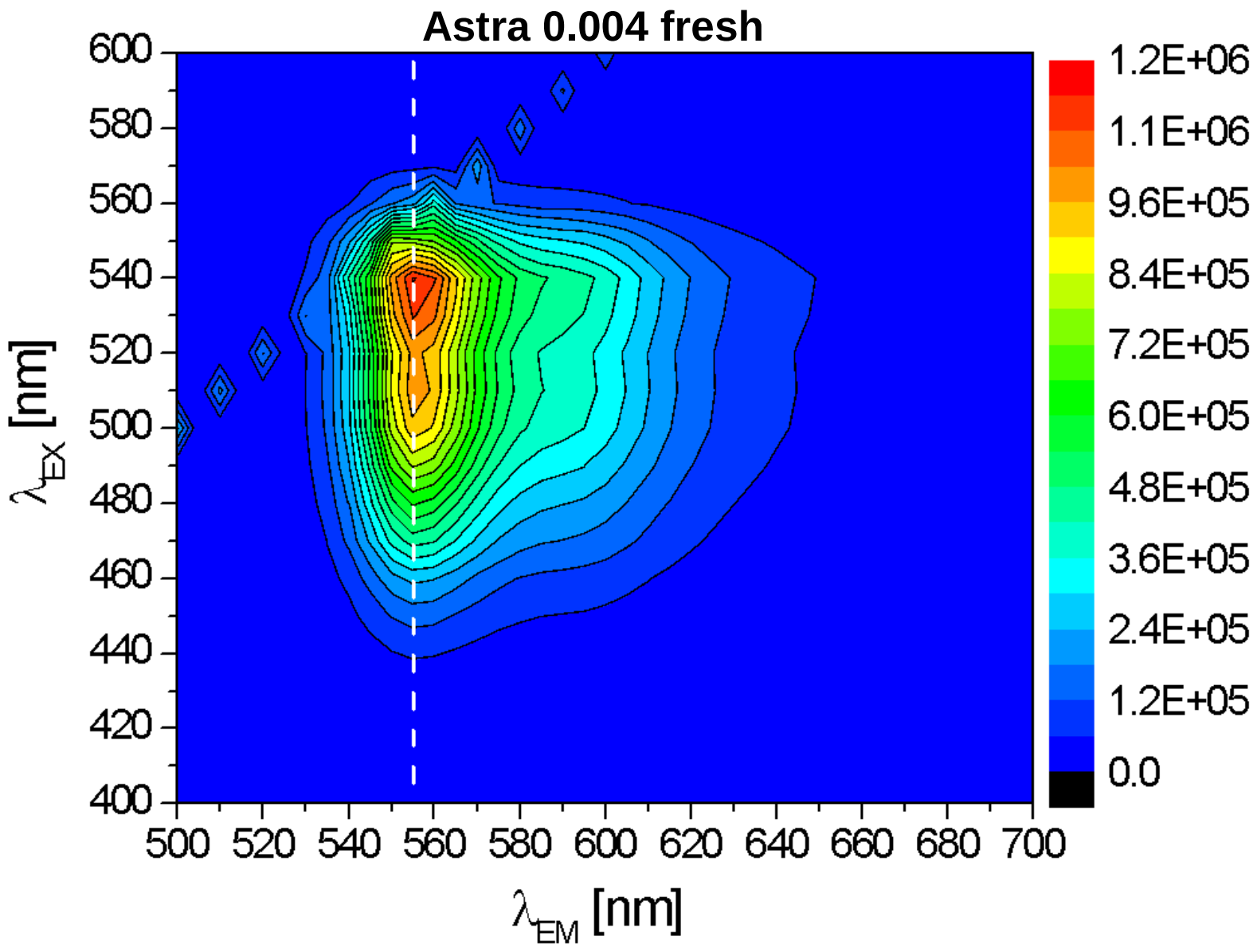

**Figure 21 – PLE maps for the dyes in visible range at concentration 0.004 mg/mL: (a) bis-astraphloxin and (b) bis-astraphloxin.**

### 4.5.2 Neat CNT and A Mixture of CNT with Bis-astraphloxin

a)

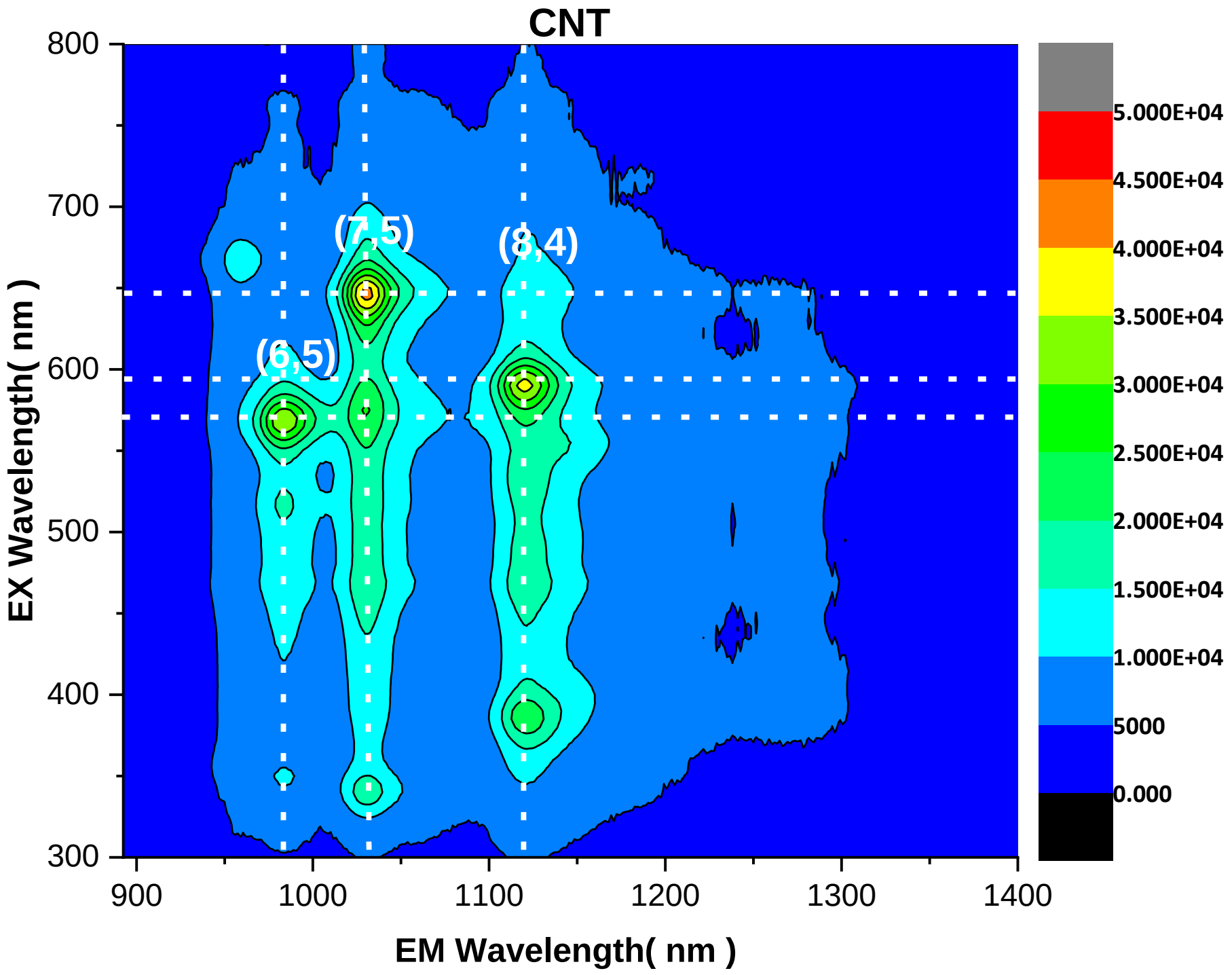

b)

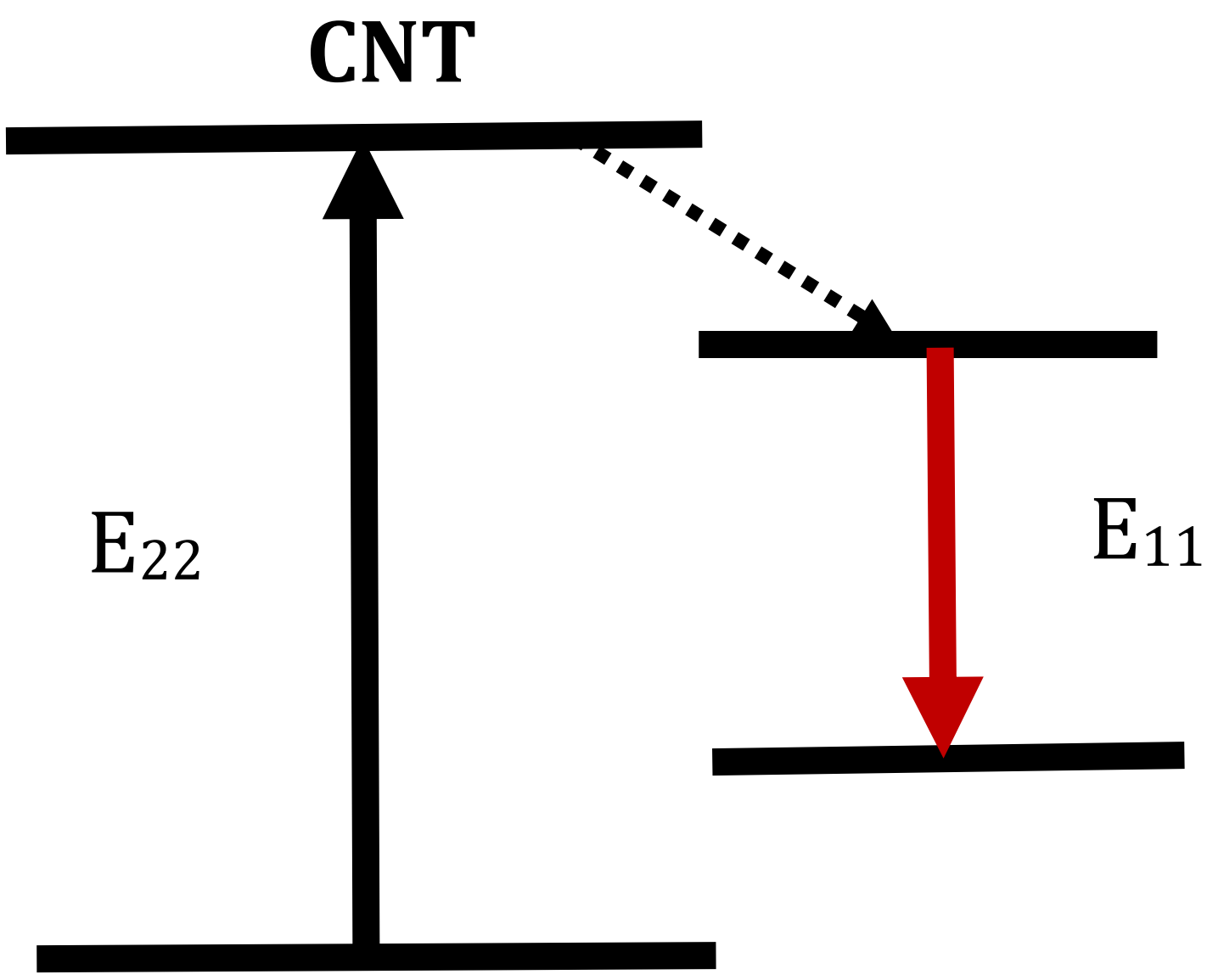

c)

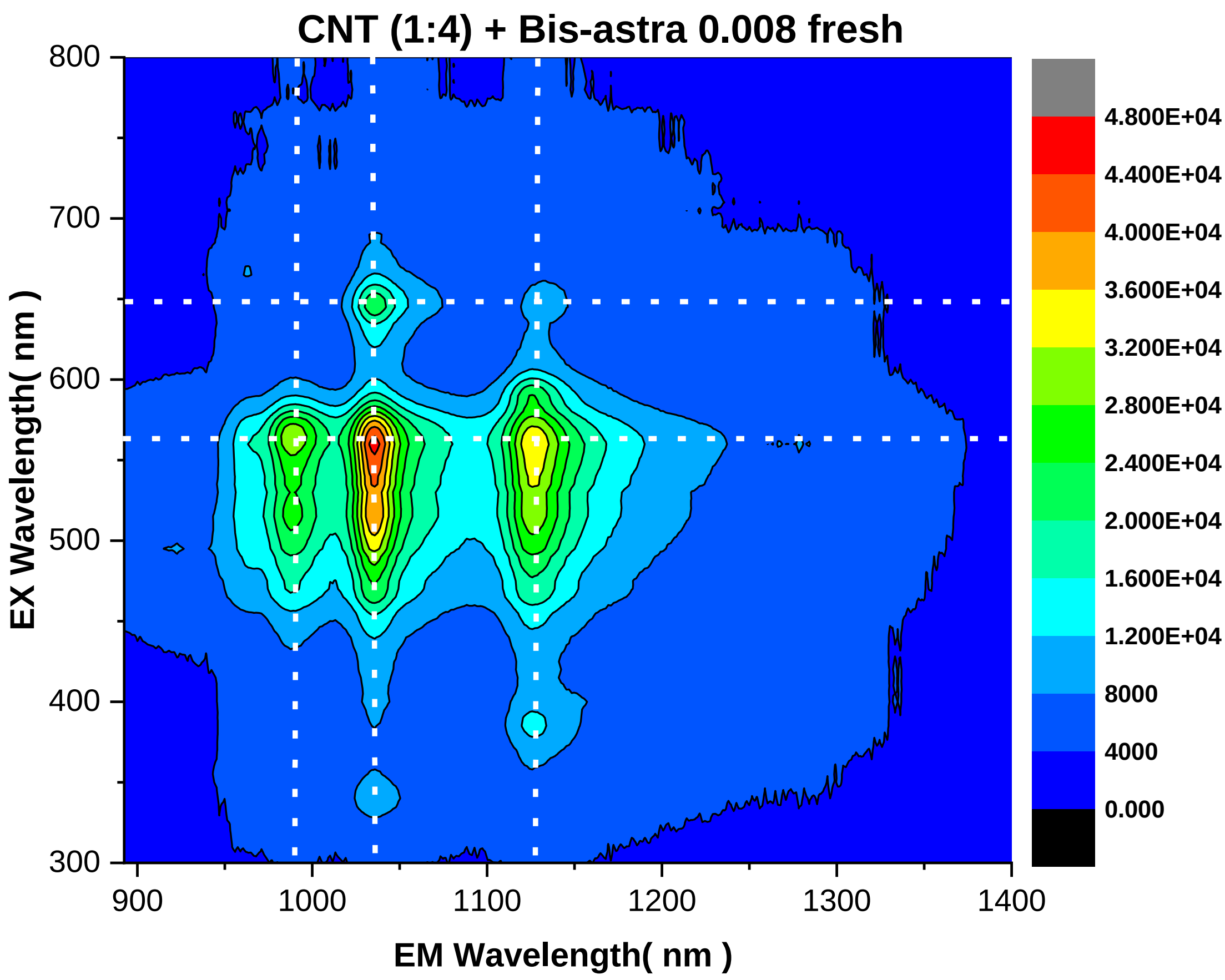

d)

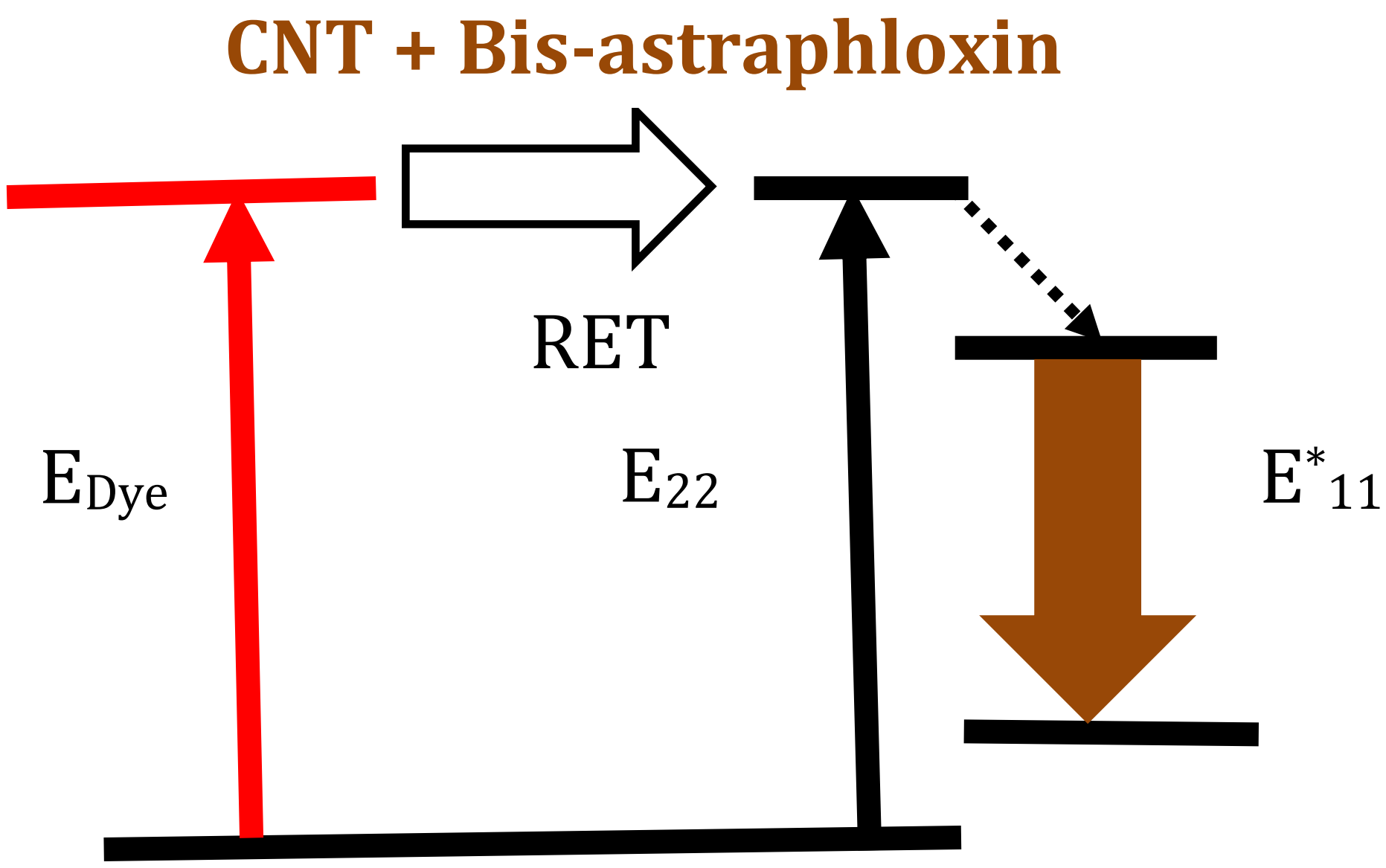

**Figure 22 – PLE map for water solutions of (a) neat CNTs dispersed with SDBS. (b) Energy diagram indicates that (a) CNTs have specific exciton energy levels of PL excitation ($E_{22}$) and emission ($E_{11}$) in the visible and NIR ranges, respectively. (c) PLE map for a mixture of CNTs with bis-astraphloxin (0.004 mg/mL) (d) a visible-range excitation of the dye molecules ($E_{Dye}$) attached to the nanotubes is transferred via RET to these CNT levels ($E_{22}$ with subsequent relaxation to $E^*_{11}$ in the NIR range).**

### 4.5.3 Mixtures of CNT with Dyes.

a)

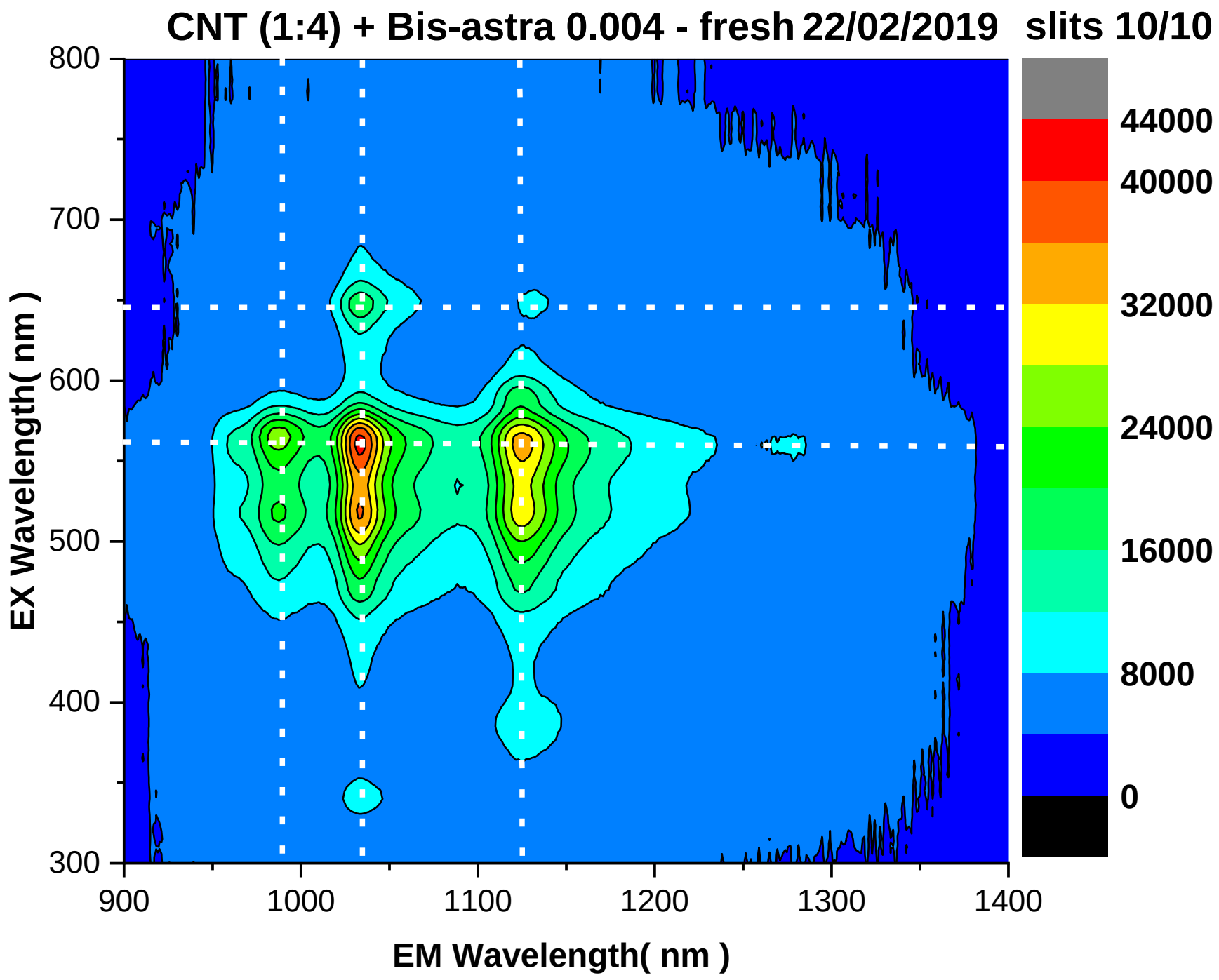

b)

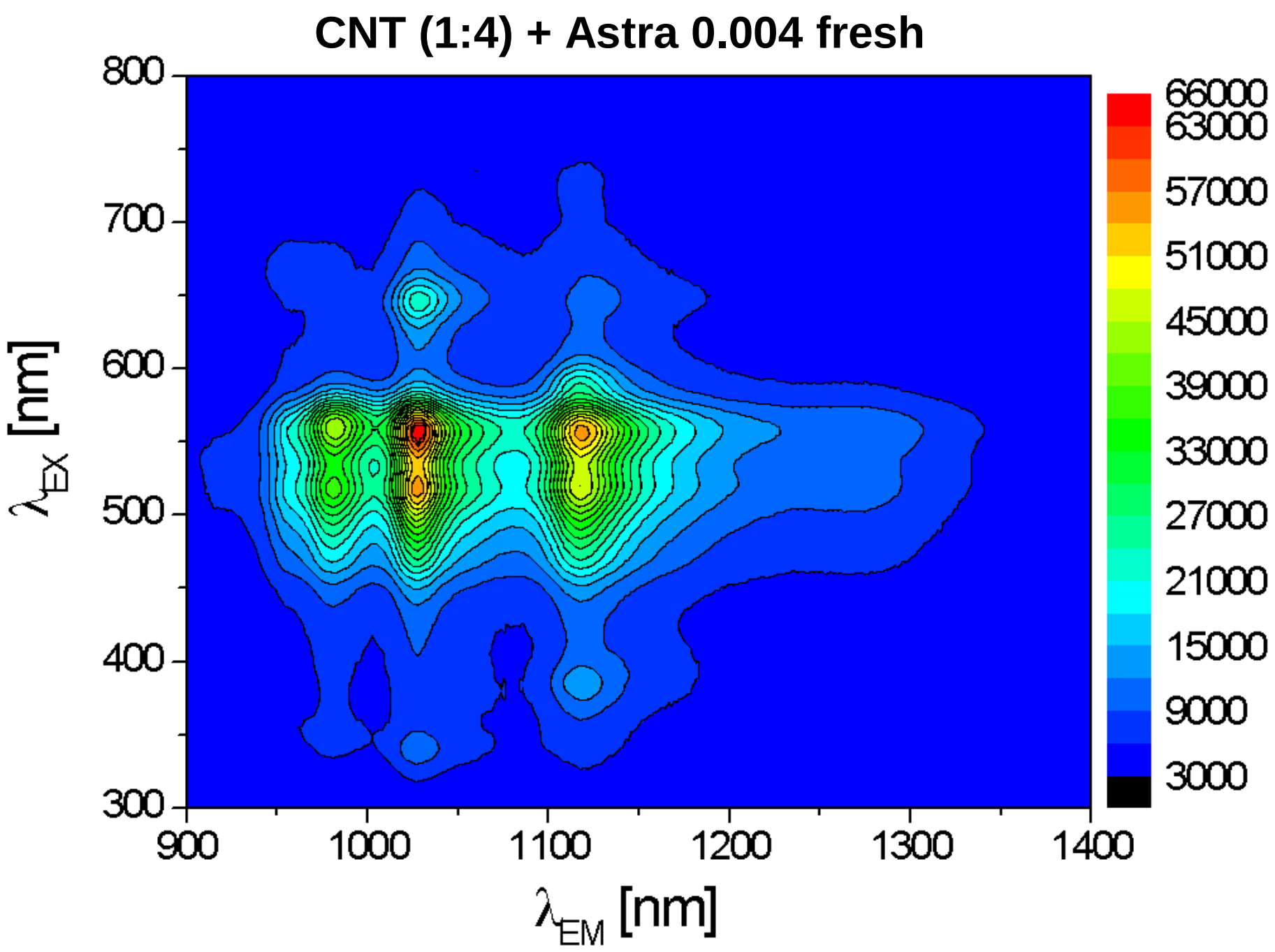

**Figure 23 – PLE maps for the mixtures of CNTs with dyes at concentration of 0.004 mg/mL: (a) bis-astraphloxin and (b) astraphloxin.**

### 4.5.4 PL Spectra of Neat CNT and the Mixtures

a)

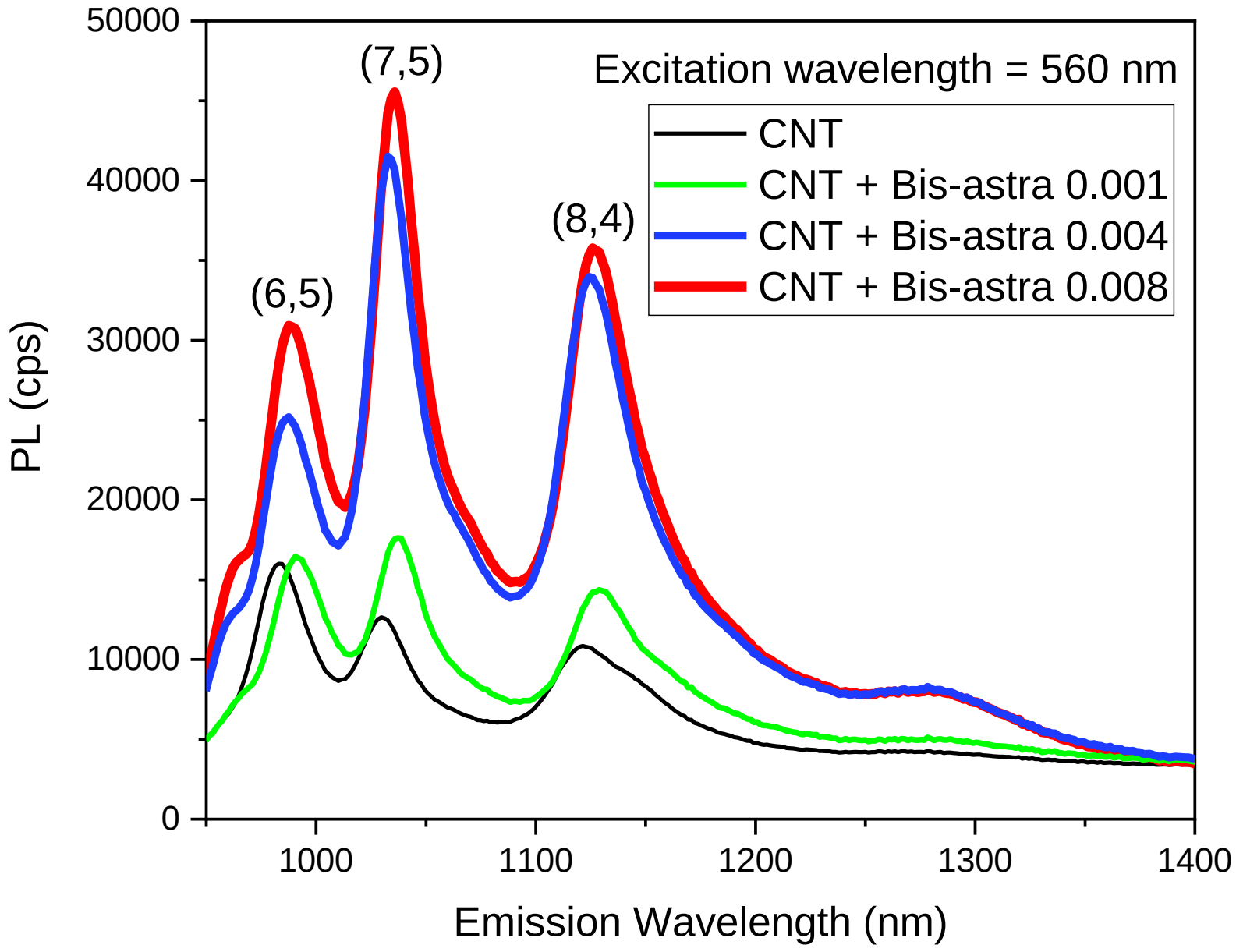

b)

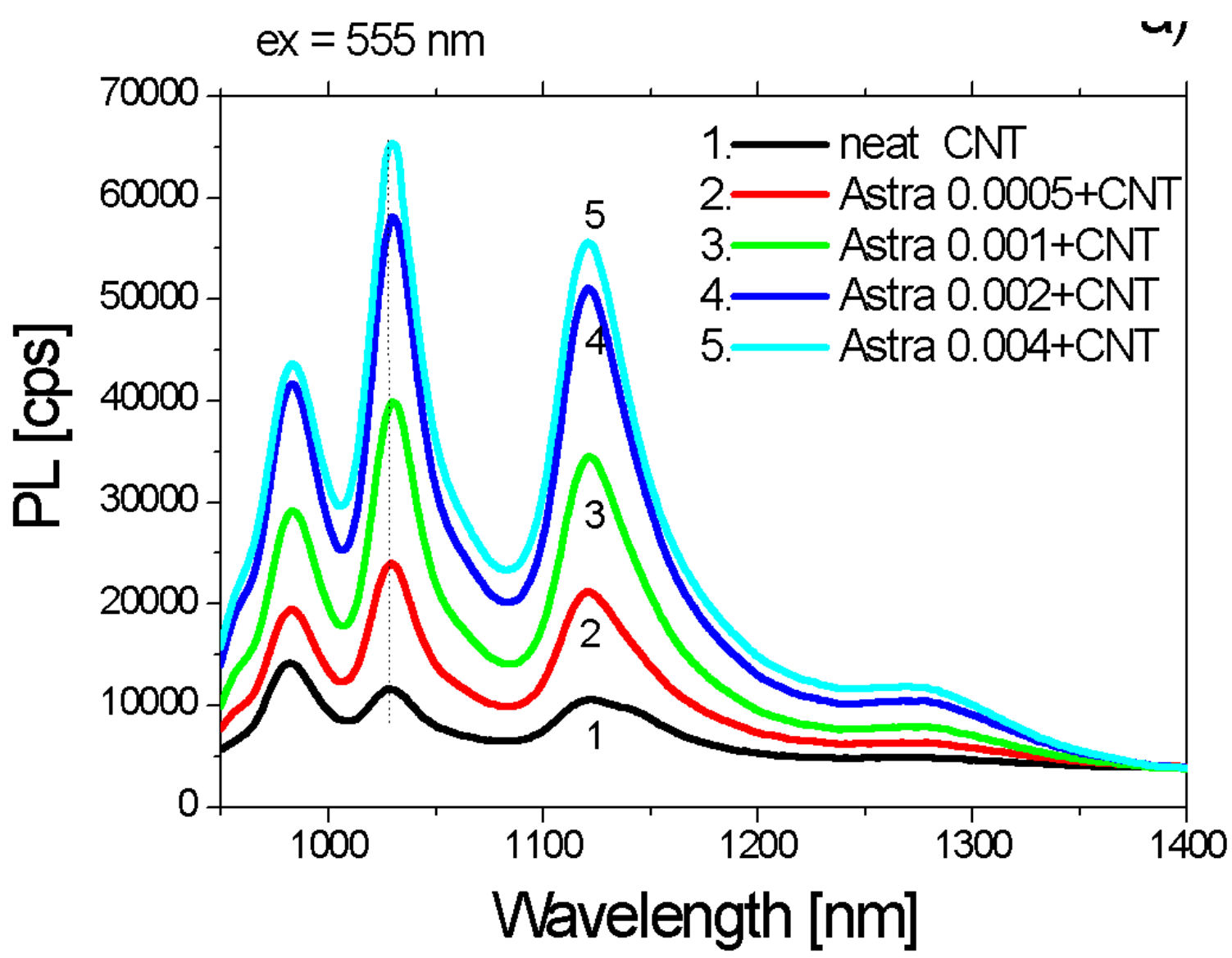

**Figure 24 – Dependence of PL on organic dye concentration in the mixtures with CNTs. (a) PL spectra of water solutions of neat CNTs and mixtures of CNTs with bis-astraphloxin (0.001 mg/mL, 0.004 mg/mL, 0.008 mg/mL). (b) PL spectra of water solutions of neat CNTs and mixtures of CNTs with astraphloxin (0.0005 mg/mL, 0.001 mg/mL, 0.002 mg/mL, 0.004 mg/mL).**

### 4.5.1 A Mixture of CNT with Bis-astraphloxin Measured Fresh and Measured After a Week

a)

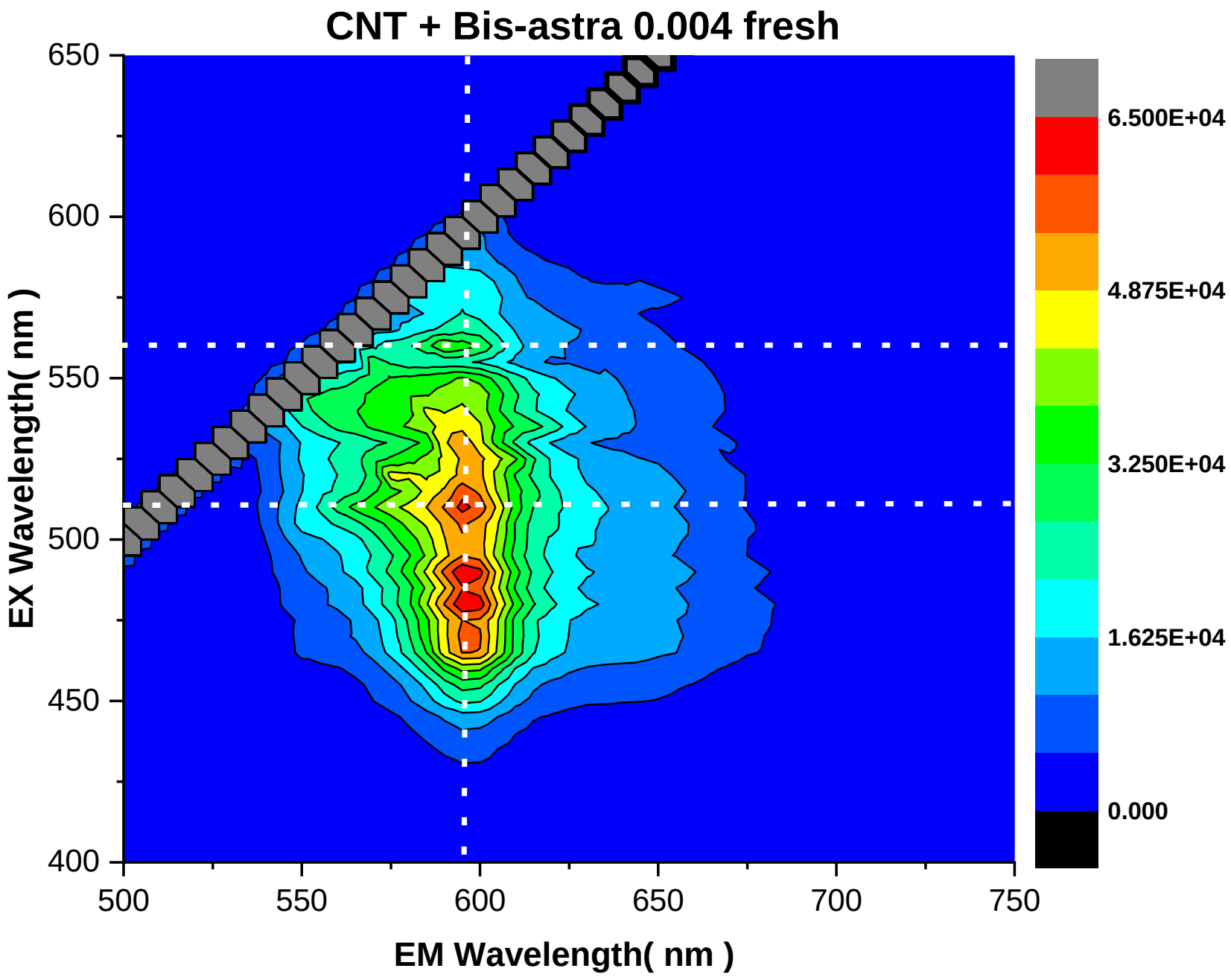

b)

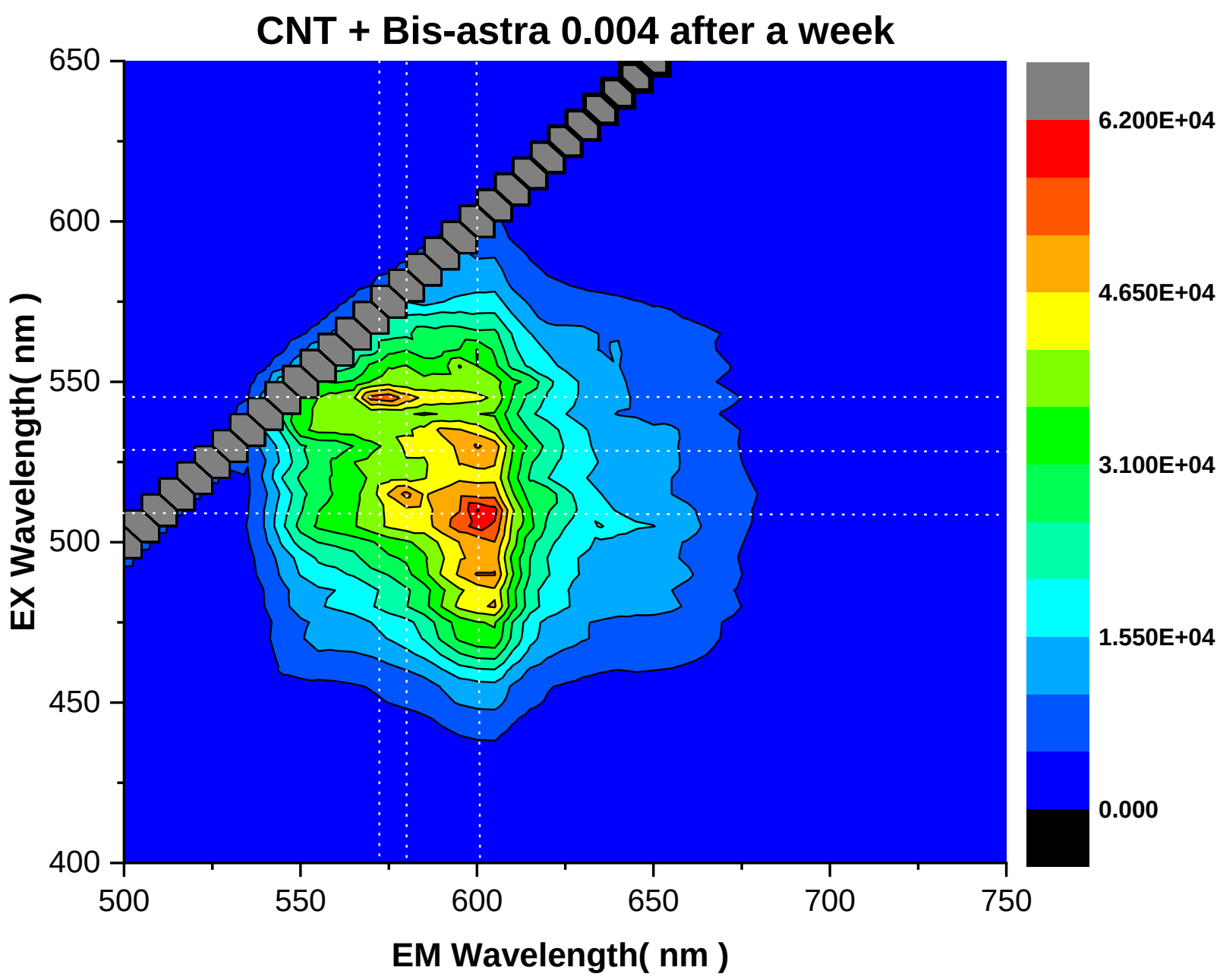

**Figure 25 – PLE maps for the mixture of CNT with bis-astraphloxin (0.004 mg/mL): (a) fresh and (b) after a week.**

## 4.6 Discussions/Analysis

In the visible range, the measured PL excitation-emission (PLE) maps for the neat bis-astraphloxin and astraphloxin (both 0.004 mg/mL) as in Figure 21 indicates a maximum intensity of $2.3 \times 10^5$ and $1.2 \times 10^6$, respectively. It should be noted, that the astraphloxin has much higher PL intensity than the bis-astraphloxin of the same concentration. For both dyes is practically the same: $\lambda_{EX}$ = 540 nm, however $\lambda_{EM}$ = 556 nm for astraphloxin and $\lambda_{EM}$ = 565 nm for bis-astraphloxin (Figure 21). Both accurate solutions of the dyes remained stable for weeks with no changes in intensity and aggregation formation.

The PL excitation-emission (PLE) maps measured in the NIR range for the neat CNTs and mixture of CNTs with bis-astraphloxin (0.008 mg/mL) are shown in Figure 22. The PLE maps for the neat CNTs (Figure 22a) with $\lambda_{EX}$ at $E_{22}$, $E_{33}$, and $E_{44}$ and $\lambda_{EM}$ at $E_{11}$ revealed the presence of CNTs with chiralities of (6,5), (7,5), (8,4) in the studied dispersions. The PLE maps of the mixtures of CNTs with bis-astraphloxin indicated a significant increase in the PL intensity from tubes of all chiralities ($E^*_{11}$) in the dye excitation range, with a maximum at $\lambda_{EX}$ = 560 nm. The peak of $\lambda_{EX}$ does not coincide with $\lambda_M$ of bis-astraphloxin, the absorption maximum of the monomers (see Figure 19a). However, it does coincide with the absorption maximum for the interaction of the dye with the CNTs at all the three concentrations levels (0.001 mg/mL, 0.004 mg/mL and 0.008 mg/mL; see Figure 19a).

Similarly, the PLE maps of the mixtures of CNTs with astraphloxin also indicated a significant improvement in the PL intensity from all tubes of all chiralities ($E^*_{11}$) in the dye excitation range, with a maximum at $\lambda_{EX}$ = 555 nm (Figure 23b). The peak of $\lambda_{EX}$ does not coincide with $\lambda_M$ as well (see Figure 19b), the absorption maximum of the monomers and mixtures at higher concentrations (0.002 and 0.004 mg/mL). Notwithstanding, it does coincide with the absorption maximum (554 nm) for the interaction of the dye with the CNTs at lower concentration (0.0005 mg/mL).

Therefore, resonant energy transfer (RET) from an excited molecule of dye (donor) attached to CNT (acceptor), a different molecule meaning hetrotransfer, is the most apparent cause of this effect and can be explained in terms of dipole-dipole interactions between the donor and acceptor. Moreover, the astraphloxin monomers present in the mixture at high concentration do not transfer energy to the CNTs, as seen from the absence of PL peaks at $\lambda_{EX} = \lambda_M$ = 540 nm (Figure 23b). Whereas, the astraphloxin molecule at lower concentration (0.0005 mg/mL) and bis-astraphloxin molecule at all the three different dye concentrations (0.001 mg/mL, 0.004 mg/mL and 0.008 mg/mL) do transfer energy to the CNT as seen from the presence of PL peaks at $\lambda_{EX} = \lambda_M$ = 555 nm and $\lambda_{EX} = \lambda_M$ = 560 nm, respectively. This means that short-range energy transfer of the Direct charge exchange (Dexter type transfer) is more likely because it occurs at a closer proximity (<1 nm) than a Förster mechanism for RET, which is sensitive at distances of 1-10 nm between donor and acceptor molecules.

For the purpose of illustrating the intensification of CNT emission due to the energy transfer, the extracted PL emission spectra at $\lambda_{EX}$ = 560 nm for the bis-astraphloxin and $\lambda_{EX}$ =555 nm for the astraphloxin from the PLE maps are shown in Figure 24(a) and Figure 24(b) respectively. The maximum PL enlargement of about 6 times was achieved at astraphloxin concentration of 0.004 mg/mL. Whereas, the maximum PL amplification was achieved at a bis-astraphloxin concentration of 0.008 mg/mL. Additionally, the PL spectra indicate the different effects of the dye on CNTs of each chirality, revealing selectively in the complexation with CNTs of various diameters (Figure 24).

According to Figure 24a, the PL intensity by the emission wavelength corresponding to a mixture of CNT with bis-astraphloxin of concentration 0.008 mg/mL and to neat CNTs of various chiralities – (6,5), (7,5), and (8,4), the highest (up to a factor of 4) PL amplification achieved for the (7,5) chirality. In addition, the PL intensity by the emission wavelength corresponding to a mixture of CNT with astraphloxin of concentration of 0.004 mg/mL and to neat CNTs of various chiralities – (6,5), (7,5) and (8,4), the highest (up to a factor of 6) PL amplification achieved for the (7,5) chirality (Figure 24b). The reason why the astraphloxin achieved a higher amplification than the bis-astraphloxin is believed to be due to more complex electronic structure of the bis-astraphloxin molecules.

Among the different CNT chiralities revealed in Figure 22a, the largest diameter or surface area corresponds to (8,4) chairality, which implies that a greater number of the dye molecules can complex with the CNTs for exhibiting the highest PL intensification, but it fails to achieve the peak PL amplification. This observation demonstrates that the electronic structure frameworks of both the CNTs and the organic dyes must simultaneously influence the PL magnification.

Comparison of the PLE maps in visible range for neat bis-astraphloxin (Figure 21a) and mixtures of the dye with CNT (Figure 25) shows aggregates formation with prominent red-shifted bands at $\lambda_{EM}$ = 600 nm. In addition, the monomer emission becomes strongly quenched in such systems. The bis-astraphloxin aggregation in CNTs is featured by one peak ($\lambda_{EM}$ = 600 nm), whereas the astraphloxin aggregation resulted in multiple peaks [16]. It evidences formation of one type of aggregates by bis-astraphloxin, and various types for astraphloxin system.

Temporal or delayed measurements of the PLE maps of the neat dye (bis-astraphloxin or astraphloxin) and neat CNT dispersions aged over a week did not show any changes in the spectra (Appendix). The mixtures of CNT-bis-astraphloxin have some significant changes over time in spectral shape. The NIR PL of mixtures with low concentration (0.001 and 0.004mg/mL) of the dye grows, when aged over a week. Oppositely, the mixture at high concentration of the dye (0.008 mg/mL) when aged over six days has lower PL intensity in NIR. Such dynamics correlate with the changes observed for absorption evidencing slightly improved complexation with time for the dye molecules at low dye the concentrations.

## 4.7 General Challenges and Mitigation

Since nanoscience field of research was a new thing to me and I was very inexperienced in this area of scientific research. It was at first a very challenging and I did not know where to start from. Therefore, after few meetings with my primary supervisor I started with a literature survey by devoting a lot of time into reviewing articles and focusing mainly on the key subjects related to this project.

Familiarizing myself with the laboratory equipment and understanding the working principles of the absorption and photoluminescence spectrometers through training was not just straightforward. Carrying out sample preparations and running measurements was very demanding. I had to put in lots of time to comprehend the experimental techniques of running sample measurements. Data analysis using OriginLab software was not an easy task, as it was a completely new platform. I spent lots of time by reading and watching videos to understand how the programme works. OriginLab turned out to be an excellent platform of choice for data analysis and graphing. It enabled an easy-to-use interface for beginners without the need for a written language. The software provided an essential set of tools for performing exploratory and advanced analysis of data.

Due to the number of people that uses the spectrometer for other research experiments, I find it very hard at times to take immediate or delayed measurements and this did not help me to achieve the time and date a certain measurement needs to be taken. Therefore, I must wait for the next day in some instances.

Finally, the fluorescence spectrometer sometimes gives inaccurate map results which could be due to technical issues or ambient light interfering with the sample. This was normally tackled by making sure that the sample compartment is properly closed with the lid and repeat the measurement which is time-demanding because each run in the NIR can take up to 1 hour 45 minutes.

Even though I encountered some of these challenges but in general, I really enjoyed this journey because it helps me to understand current experimental techniques for scientific spectral characterization of materials by absorption, photoluminescence excitation and emission mapping. The project also shows how spectroscopy science is helping to identify materials to characterize material interactions in various environments, and to design new compounds.

# 5 Conclusion

In this dissertation, experiments conducted in the visible and near-infrared range are geared towards understanding energy transfer in the system of SWCNT-surfactant-dye and for characterizing SWCNTs using photoluminescence spectroscopy.

The first primary objective was to do a literature review on SWCNTs functionalization, energy transfer, and photoluminescence imaging as a chapter was carried out and completed transforming lately into Chapter 2 (Background).

The second primary objective was to perform experimental work for mixtures of SWCNT with bis-astraphloxin. The journey started by going through some training on experimental measurements of absorption and photoluminescence spectra, including excitation-emission photoluminescence maps. Followed by samples preparation: neat dispersion of SWCNT, neat solution of the dye, and mixtures of the SWCNT with bis-astraphloxin and then by conducting immediate measurements of absorption and photoluminescence spectra for the samples. The absorption and photoluminescence spectra for mixtures of SWCNT with bis-astraphloxin have been measured. Besides, as a reference, the spectra for the mixture of SWCNT with astraphloxin have also been recorded.

The final primary objective is focused on investigation of the physical mechanism for the energy transfer in the SWCNT-dye system via measured spectral data. It was shown that SWCNT interacts with bis-astraphloxin as the spectra of the mixtures are not a superposition of the spectra of each component: SWCNT or bis-astraphloxin dye only. In the studied mixtures of SWCNT with bis-astraphloxin, it was observed a strong quenching of visible range dye emission and significant enhancement (up to a factor of 4) of the NIR emission of SWCNT in the range of the dye excitation. Plotting the PL intensity vs the emission wavelength for corresponding to a mixture of CNT with astraphloxin of concentration of 0.004 mg/mL and to neat CNTs of various chiralities – (6,5), (7,5) and (8,4), the highest (up to a factor of 6) PL amplification achieved for the (7,5) chirality. This is evidencing efficient energy transfer from the interacting bis-astraphloxin (donor), to SWCNT (acceptor). Comparison of the above mixture with the mixture of astraphloxin-SWCNT and relating to literature data [2], it is possible to associate the interaction and energy transfer due to the formation of ionic complexes of bis-astraphloxin-SWCNT. The complexes are self-assembled due to Coulomb attraction of positively charged bis-astraphloxin and positively charged SWCNT covered by negatively charged surfactant.

The extended objective explored by experimenting and analysing of temporal developments of energy transfer/aggregation in the SWCNT-dye mixtures. Delayed measurements of absorption and photoluminescence spectra for the samples are carried out for updated physical mechanism for the SWCNT-dye energy transfer. It was found that neat solutions of the dye remained stable for weeks with no aggregation formation. The temporal studies show that energy transfer complexes bis-astraphloxin-SWCNT are formed immediately, however their properties were slightly changing (over a week) depending on the concentration of the dye in the mixture. At low concentration of dye (0.001g/L), the energy transfer peaks of photoluminescence were growing, and for high concentrations (0.008g/L) the energy transfer peaks of emission had a bit lower intensity.

# 7 Appendix- MAPS

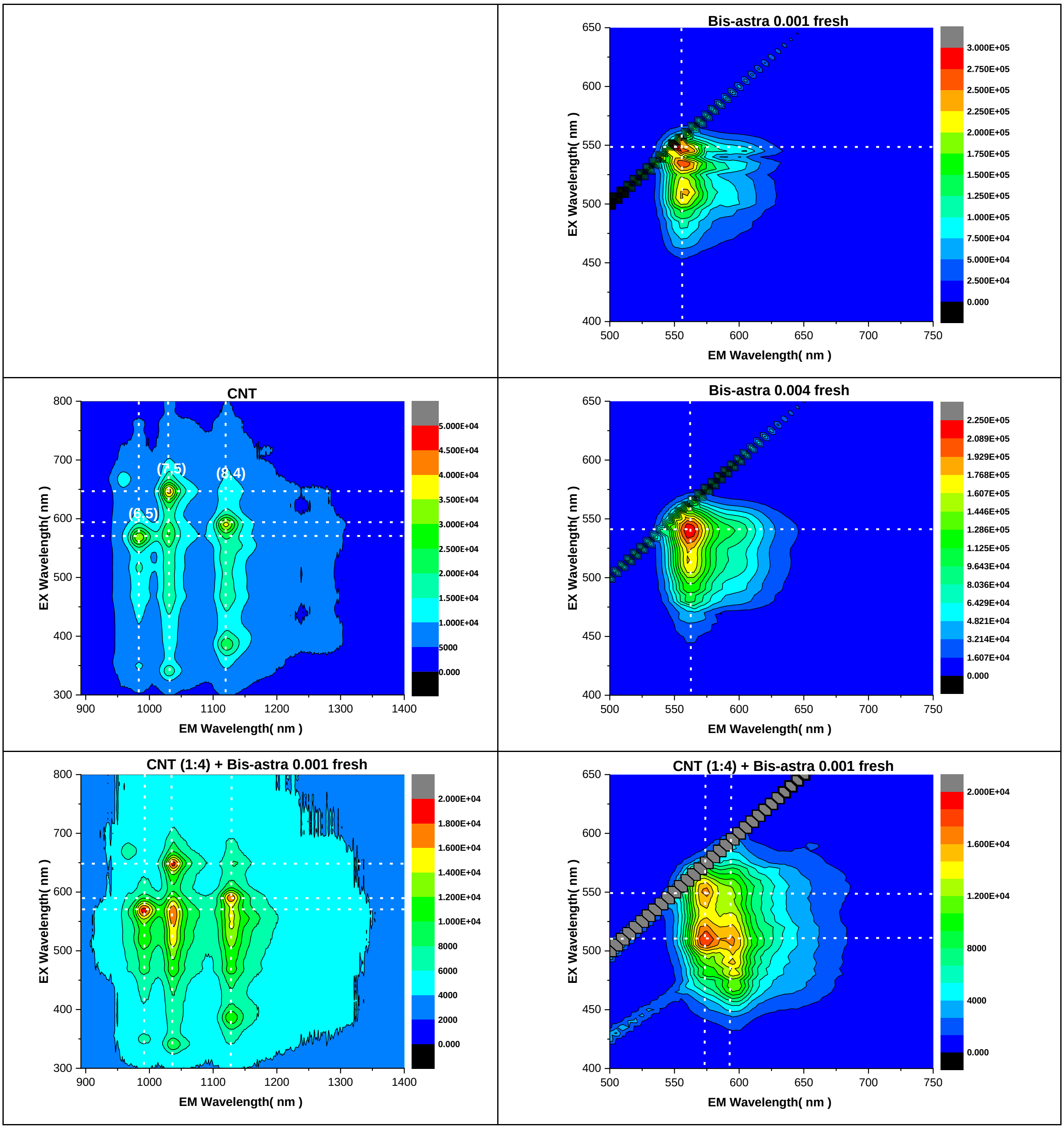

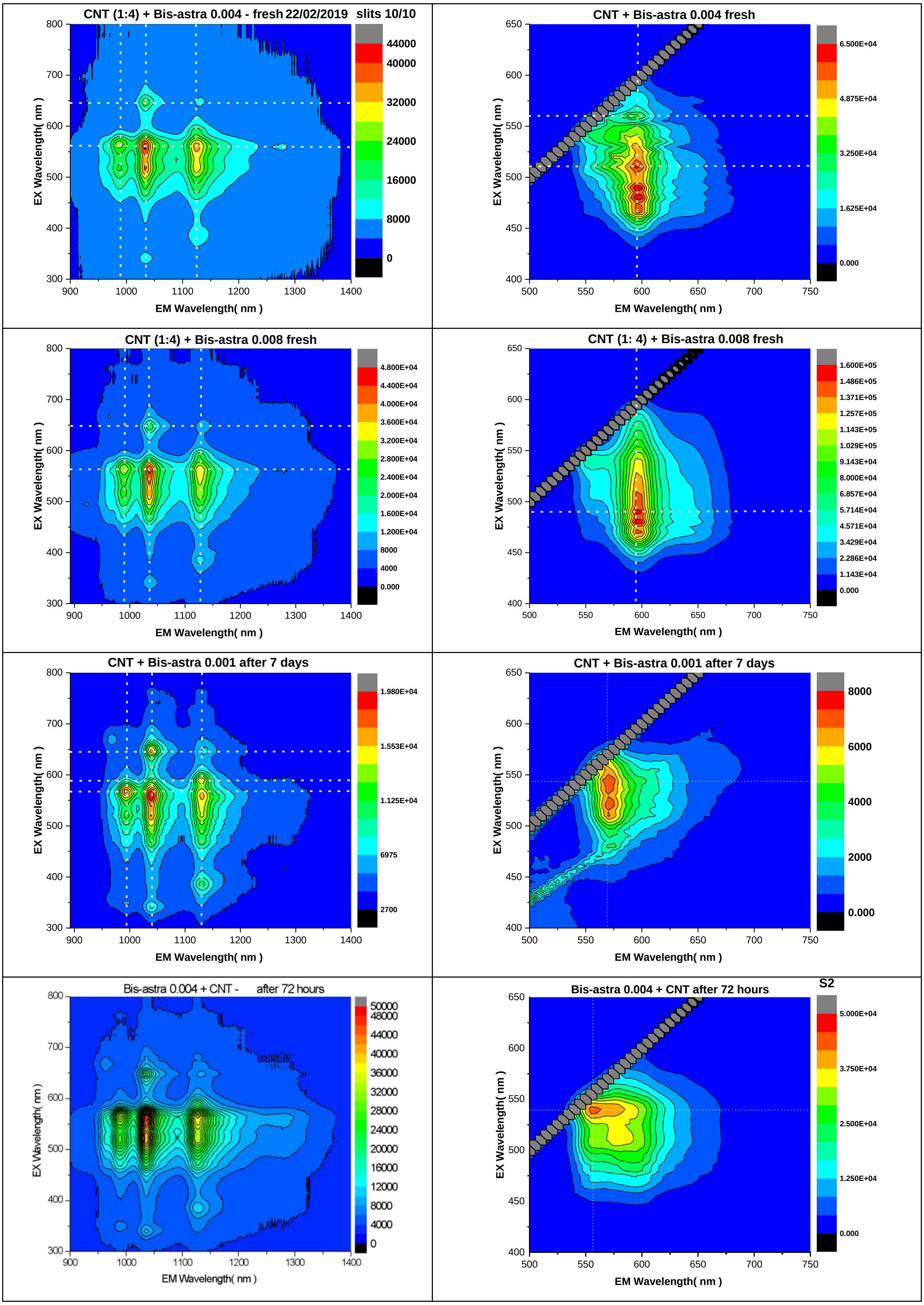
CNT (1:4) + Bis-astra 0.004 - fresh 22/02/2019 slits 10/10
EX Wavelength( nm )
EM Wavelength( nm )
44000
40000
32000
24000
16000
8000
0
CNT + Bis-astra 0.004 fresh
6.500E+04
4.875E+04
3.250E+04
1.625E+04
0.000
CNT (1:4) + Bis-astra 0.008 fresh
4.800E+04
4.400E+04
4.000E+04
3.600E+04
3.200E+04
2.800E+04
2.400E+04
2.000E+04
1.600E+04
1.200E+04
8000
4000
0.000
CNT (1: 4) + Bis-astra 0.008 fresh
1.600E+05
1.486E+05
1.371E+05
1.257E+05
1.143E+05
1.029E+05
9.143E+04
8.000E+04
6.857E+04
5.714E+04
4.571E+04
3.429E+04
2.286E+04
1.143E+04
0.000
CNT + Bis-astra 0.001 after 7 days
1.980E+04
1.553E+04
1.125E+04
6975
2700
CNT + Bis-astra 0.001 after 7 days
8000
6000
4000
2000
0.000
Bis-astra 0.004 + CNT - after 72 hours
50000
48000
44000
40000
36000
32000
28000
24000
20000
16000
12000
8000
4000
0
S2
Bis-astra 0.004 + CNT after 72 hours
5.000E+04
3.750E+04
2.500E+04
1.250E+04
0.000

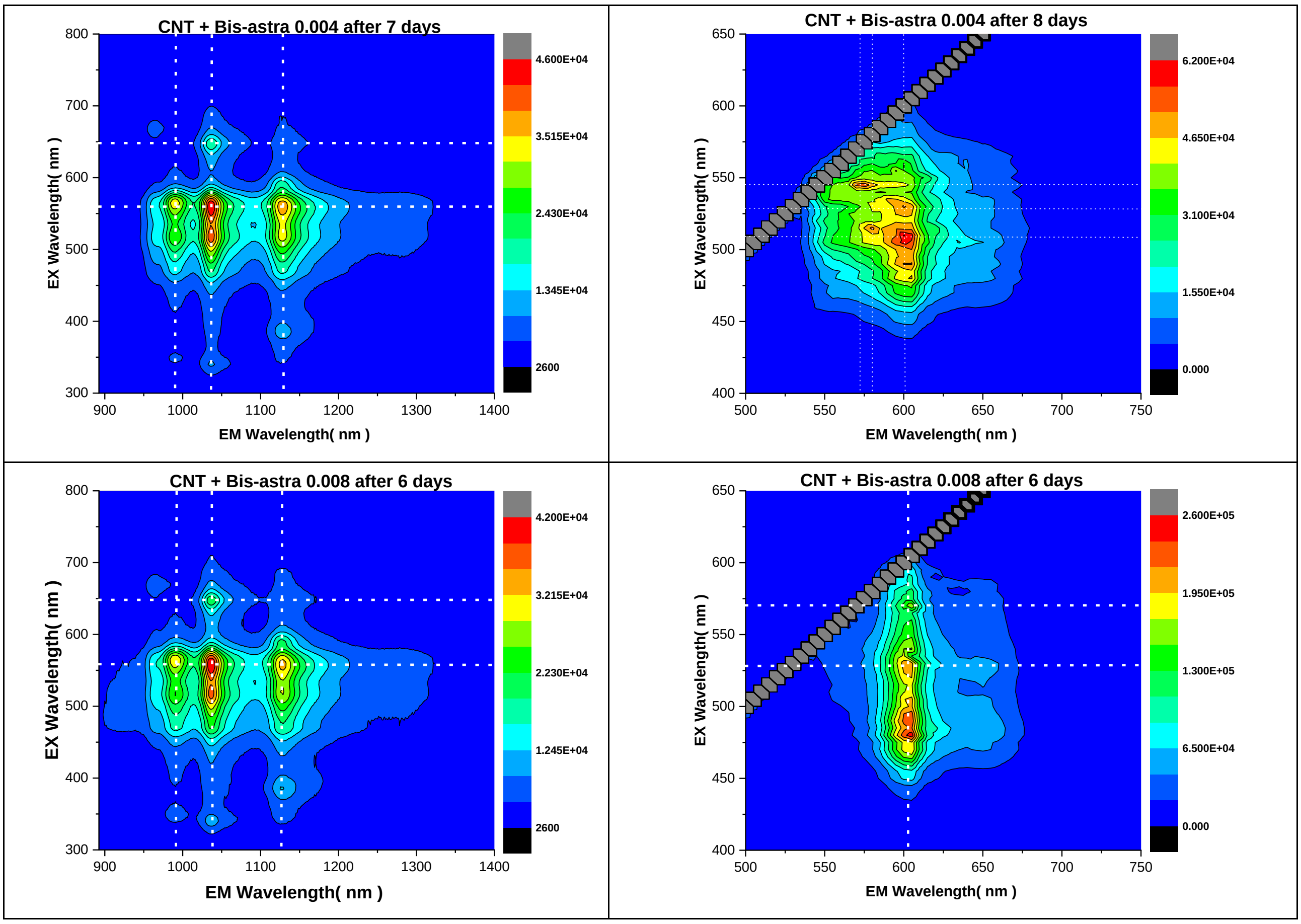
CNT + Bis-astra 0.004 after 7 days
EX Wavelength( nm )
EM Wavelength( nm )
4.600E+04
3.515E+04
2.430E+04
1.345E+04
2600
CNT + Bis-astra 0.004 after 8 days
EX Wavelength( nm )
EM Wavelength( nm )
6.200E+04
4.650E+04
3.100E+04
1.550E+04
0.000
CNT + Bis-astra 0.008 after 6 days
EX Wavelength( nm )
EM Wavelength( nm )
4.200E+04
3.215E+04
2.230E+04
1.245E+04
2600
CNT + Bis-astra 0.008 after 6 days
EX Wavelength( nm )
EM Wavelength( nm )
2.600E+05
1.950E+05
1.300E+05
6.500E+04
0.000